\documentclass[sort&compress]{elsarticle}

\usepackage{amsmath}
\usepackage{amssymb}
\usepackage{mathrsfs}

\journal{Physics of Life Reviews}

\begin{document}

\begin{frontmatter}



\title{The Emergence of Modularity in Biological Systems}


\author[phy]{Dirk M. Lorenz}
\author[ok]{Alice Jeng}

\author[phy]{Michael W. Deem \corref{cor1}}
\ead{mwdeem@rice.edu}

\address[phy]{Department of Physics and Astronomy, Rice University}
\address[ok]{Department of Chemical, Biological, and Materials Engineering, University of Oklahoma}

\cortext[cor1]{Corresponding author}

\begin{abstract}
In this review, we discuss modularity and hierarchy
in biological systems.  We review examples from
protein structure, genetics, and biological networks
of modular partitioning of the geometry of biological space.
We review theories to explain modular organization of biology, with
a focus on  explaining how biology may spontaneously organize
to a structured form.  That is, we seek to explain how
biology nucleated from among the many possibilities in chemistry.
The emergence of modular 
organization of biological structure will be described
as a symmetry-breaking phase transition, with modularity as
the order parameter.
Experimental support for this description will be reviewed.
Examples will be presented from pathogen structure, metabolic
networks, gene networks, and protein-protein interaction networks.
Additional examples will be presented from ecological food networks,
developmental pathways, physiology, and social networks.

\end{abstract}


\end{frontmatter}



\begin{quotation}
There once were two watchmakers, named Hora and Tempus, who manufactured very fine watches. Both of them were highly regarded, and the phones in their workshops rang frequently --- new customers were constantly calling them. However, Hora prospered, while Tempus became poorer and poorer and finally lost his shop. What was the reason?

The watches the men made consisted of about 1,000 parts each. Tempus had so constructed his that if he had one partly assembled and had to put it down --- to answer the phone say --- it immediately fell to pieces and had to be reassembled from the elements. The better the customers liked his watches, the more they phoned him, the more difficult it became for him to find enough uninterrupted time to finish a watch.

The watches that Hora made were no less complex than those of Tempus. But he had designed them so that he could put together subassemblies of about ten elements each. Ten of these subassemblies, again, could be put together into a larger subassembly; and a system of ten of the latter sub-assemblies constituted the whole watch. Hence, when Hora had to put down a partly assembled watch in order to answer the phone, he lost only a small part of his work, and he assembled his watches in only a fraction of the man-hours it took Tempus.''
\end{quotation}
\begin{flushright}
H.\ A.\ Simon, The Architecture of Complexity, 1962 \cite{Simon1962}.
\end{flushright}

\section{Introduction}
As Simon's classic parable of Hora and Tempus illustrates, there are advantages to assembling a complex system from modular pieces in a hierarchical fashion. This is particularly true in evolutionary adaptation, where a modular structure provides many benefits. The space of all genotypes is exponentially large, making an exhaustive search for fitness maxima impossible even on evolutionary time scales. But a system that can be decomposed into modules can be evolved one module at a time.  Thus, modularity can reduce the task of searching the entire space of possibilities into a polynomial problem of searching in the subspace of modular solutions. A physicist might recognize the similarity to a separable Hamiltonian while a geneticist might describe this
decomposition as a reduction of pleiotropic effects. Separating a complex system into independent components allows for the separate evolution of each component. Modules may change with limited perturbation to other modules. In addition, once these modules exist, new functions can be generated by combinatorial recombination of these modules rather than invention of new functionality from scratch.

Modularity is an important property in biology
because it helps a system `save its work' while allowing further evolution. In the natural world, one often finds modular, hierarchical structures. In this article, we will discuss the conditions under which 
formation of such structure may be thought of  as a symmetry-breaking phase
 transition.  We will discuss how modularity gives biological
 systems a greater ability to respond to change.  
We will review how modular structure might form spontaneously.

Modularity provides biology with a basis set to explore the space
of biological possibility.
 From a computer science point of view, an evolving system may approximate the NP-hard problem of searching all of configuration space with a polynomial-hard problem by becoming modular and hierarchical.  The analogy in physics would be achieving separability of a glassy Hamiltonian, or block diagonalization in quantum chemistry.  The modular subproblems are much easier to solve, and the partial solutions are efficiently recombined to find solutions to the original problem.  

What is the drawback?  The drawback is that the system has placed a constraint upon the space of states to be considered.  The modular and hierarchical subspace is exponentially smaller than the original space.  This is the reason for the NP $\to$ P transition. Thus, there is a trade off between increased speed of convergence toward greater fitness versus a reduced density of states of solutions in a modular framework.  

The advantage of modularity is commonly employed in engineering,
leading
some to encourage the use of modularity in evolutionary design \cite{Lipson2002}. Biological systems, however, are not designed; they are shaped by evolution. Explaining the evolutionary emergence of modularity has been a challenge, and so far no consensus has been reached \cite{Espinosa-Soto2010}. Often in biology modularity is presumed to exist \emph{a priori}, \emph{e.g.}\ the genome can be decomposed into genes, even though it has been recognized that modular solutions make up only a tiny fraction of the solution space and often optimal solutions are not modular \cite{Kashtan2009a}. In other words, most functions that biological systems perform could be performed better by less modular approaches.

In our review of
the empirical evidence, we will show that natural and man-made systems employ modularity to a non-zero extent.
That is, we will show that the polynomial approximation achieved by modularity and hierarchy has evolved in real networks.  
Modularity has been observed in all parts of biology on scales from proteins and genes \cite{Litvin2009} to cells \cite{Hartwell1999, Wagner1996} to organs \cite{Schlosser2004} to ecosystems \cite{Krause2003, Montoya2006}. Proteins are often made up of almost independent modules, which may be exchanged through evolution. Pieces of DNA that encode these distinct protein modules have become organized and concatenated in the course of evolution \cite{Baltimore2001}. Topological analysis of networks of genes or proteins has revealed modularity as well. Motifs \cite{Shen-Orr2002} and modules \cite{Singh2008} have been found in transcriptional regulation networks, and modules have been found across all scales in metabolic networks \cite{Spirin2006}. Animal body plans can also be decomposed into clear structural or functional units \cite{Klingenberg2008, He2010}. Food webs also show compartmentalization \cite{Krause2003}. Thus, a hierarchy of modules can be observed that spans many scales of biology.

While most biologists agree on the existence of modularity, many different definitions are in use \cite{Callebaut2005}. A systems biologist might describe modules from a graph-theoretical point of view as groups of nodes that are more strongly intraconnected than interconnected \cite{Espinosa-Soto2010}, a geneticist might consider a set of co-expressed or co-regulated genes a module \cite{Eisen1998, Litvin2009}, and an evolutionary biologist might look for conserved sequences or structures \cite{Karlin2003}.

Many theories have been proposed to explain how and under which conditions modularity emerges. Some of the theories argue that selection is not essential for modularity \cite{Sole2008} while others have explained the emergence of modularity through direct or indirect fitness benefits such as enhanced evolvability \cite{Bogarad1999a}, facilitated horizontal gene transfer \cite{Rainey2004, McAdams2004}, or improved robustness \cite{Kitano2004}. We hypothesize that a changing environment selects for adaptable frameworks, and that competition among different evolutionary frameworks leads to selection of structures with the most efficient dynamics, which are the modular ones.
 Here we review theories for the emergence of modularity and provide empirical evidence for its emergence
under conditions for which modularity is expected to arise spontaneously.  

This article is organized as follows.
In section 2 we  review different theoretical descriptions for
how modularity might emerge.  The section culminates with a theory
for the spontaneous emergence of modularity based upon three axioms:
a changing environment, exchange of genetic material between individuals,
and a rugged fitness landscape.
In section 3 we review experimental observations of modularity.
We give examples of modular systems in pathogens, metabolic and
genetic networks,  and protein-protein interactions.
We discuss modularity in ecological networks, physiology, and
social networks.  In each case, we emphasize the spontaneous nature
of the emergence of modularity, and how environmental pressures
have lead to increasingly modular systems.
We conclude in section 4.

\section{Theoretical Models}
\subsection{Neutral Models}
Neutral theory is a base case, a null model, of evolutionary theory.
In neutral theories, evolution is considered to be an unbiased
random walk through the state space of all possibilities.  While
real evolution is clearly shaped by selection, and so evolutionary
trajectories are biased, neutral theory remains a bastion of
theoretical effort.

\subsubsection{Duplication-Differentiation}
Most neutral theories for the emergence of modularity have focused on the idea of duplication. If parts of a system are duplicated, the result will be more modular than the original system. It has been shown,
for example, that artificial networks created by a duplication operator can have a hierarchical modular structure that is similar to the structure observed in the yeast protein-protein interaction network \cite{Hallinan2004a}. A parameter of this duplication-differentiation process controls a phase transition between a highly-connected graph and a sparsely connected graph. Close to the critical value, the resulting networks are scale-free, small world network, with a modular structure \cite{Sole2008}. Sol\'e and Fernandez suggested that natural selection might have tuned this parameter so that networks are sparse but completely connected. Modularity would then emerge as a byproduct without any selection pressure. It was also shown that the distribution of subgraphs in the resulting networks matches that observed in the human interactome, the yeast proteome, and a subset of human transcription factors \cite{Sole2008}.

While the previous studies explored the purely topological growth of networks without regard for biological function, Soyer extended the idea of duplication-differentiation to include a constant selective pressure \cite{Soyer2007}. In his model, the requirement that a pathway be able to respond to two different signals leads to the emergence of modularity in regulatory pathways. However, this modularity is hard to maintain because of drift to non-modular systems with equal fitness. Including horizontal gene transfer into the model may affect the results. 

A challenge to theories for the emergence of modularity based on gene duplication comes from an empirical study \cite{Price2008} of the evolution of transcriptional regulation in \emph{E. Coli}. There, it was shown that most of the transcription factors in \emph{E. Coli} did not evolve by gene duplication but rather by horizontal gene transfer. The same study also observed two trends in the evolution of gene regulatory networks that seem inconsistent with nearly neutral theories. Many similarities between paralogs can be shown to be the result of convergent evolution. They have not been conserved from the common ancestor. In addition, the regulation of genes that were horizontally transferred tend to be more complex than that of native genes. These results cannot be explained by neutral theories.

\subsubsection{Neutral Modular Restructuring}
Force \emph{et al}.\ have developed a near-neutral model for the emergence of genotypic modularity based on mutation, duplication, and genetic drift \cite{Force2005}. In a first step, pleiotropic constraints may be reduced by neutral changes in gene architecture without altering the phenotype. This can happen if functions that were regulated together evolve to be regulated independently. The benefits of this restructuring in regulation may not be realized immediately, especially in a constant environment. However, if the environment changes, this genotypic modularity may provide a selective advantage which may promote phenotypic modularity. Force \emph{et al.}\ stress the distinction between the neutral change of the genomic architecture and its effects on the subsequent phenotypic evolution.

Neutral models may provide a theory for the initial appearance of modular structures, but they cannot explain why such a structure would persist in the presence of more optimal nonmodular structures \cite{Kashtan2005}.

\subsection{Models Involving Natural Selection}
\label{sec:smoothLandscape}
Modularity contributes positively to fitness by several indirect means. Modular systems are more robust because the effect of perturbations can be contained within a module. A failure of one part does not affect the entire system. Modularity also enhances evolvability because it allows different parts to be optimized separately without impairing the functioning of other parts. In addition, once modules exist, they can be reused to facilitate further evolutionary adaptation. New functionality does not have to be created from scratch, but rather can result from different combination of existing modules. Furthermore, a rewiring of modules can be achieved quickly in response to environmental perturbation. Finally, modularity makes the exchange of genetic information much easier. Because of these benefits, modularity speeds up evolution and can thus be selected for directly or indirectly.

Studies on a smooth fitness landscape, however, have failed to capture these benefits of modularity. Orr studied the evolution on a smooth landscape given by the Fisher model and found that the rate of adaptation decreases with increasing complexity, as measured by the dimensionality of the adaptive landscape \cite{Orr2000}. He derived analytically that the rate of change of fitness is given by
\begin{align}
    \frac{d\bar w}{dt} = -\frac{4N\mu r^2}{n} M \bar w \ln \bar w,
\end{align}
or, if time is measured in units of $(N\mu)^{-1}$ generations,
\begin{align}
    \frac{d\bar w}{dt} = -\frac{4 r^2}{n} M \bar w \ln \bar w,
\end{align}
where
\begin{align}
    M = \frac{1}{\sqrt{2\pi}} \int_x^\infty \left(y - x\right)^2 \exp\left[-\frac{y^2}{2}\right]
\end{align}
and
\begin{align}
    x = \frac{r \sqrt n}{2\sqrt{-2\ln \bar w}}.
\end{align}
Here, $\bar w$ is the average fitness, $N$ is the population size, $\mu$ is the mutation rate, $n$ is the number of independent characters, $r$ is the size of mutations in phenotype-space, $x = r \sqrt{n}/2z$ is a normalized measure of the size of mutations, and $z$ is the initial distance from the optimum. This result shows that on a smooth fitness landscape described by Fisher's model, the rate of adaptation decreases with increasing complexity at least as fast as $n^{-1}$. This is a consequence of the fact that in complex organisms random mutations are less likely to be favorable, the probability of fixation is lower, and the increase in fitness is smaller in the event of fixation; this assumes that one considers mutations whose size is independent of complexity \cite{Orr2000}. Thus, there appears to be a selective advantage in reducing complexity as measured by the number of independent characters.

Fisher's original model has a universal pleiotropy. Welch and Waxman investigated \cite{Welch2003} how modularity, which they modeled by a modularly parceled pleiotropy, would affect Orr's results on the cost of complexity. Surprisingly they found that modular pleiotropy cannot eliminate the cost of complexity. The $n^{-1}$ dependence observed by Orr is also observed if the degree of pleiotropy is restricted in a modular fashion. Modularity, as considered in their paper, increases the rate of adaptation if a single trait is maladapted, but may decrease the adaptation rate if all traits are equally maladapted. Therefore, modularity understood as a reduction in the number of traits that can be affected by a single mutation, does not necessarily provide a fitness advantage and cannot explain the increase in complexity observed over evolutionary time scales. On a smooth fitness landscape there always seems to be a cost associated with increasing complexity. Explanations for the emergence of modularity thus may have to involve a different set of assumptions than the Fischer model. For example, on a rugged fitness landscape complexity may increase the rate of evolution.

\subsubsection{Selection for Stability or Robustness}
Robustness is a generic property of biological systems \cite{Kitano2004}. It describes their ability to resist perturbations. A modular structure enhances the robustness of a system by decreasing the spread of a perturbation. Since robustness improves the fitness of an organism, and modularity contributes to fitness, modularity can be co-selected. This hypothesis has been confirmed in a study \cite{Variano2004} of linear dynamics on a network. In this model evolving individuals are represented by matrices whose fitness is defined by the number of eigenvalues whose real part is negative. As the individuals evolve towards higher fitness they develop a hierarchical modularity which improves their robustness. Modularity is implicitly selected because of the selection for robustness. 

\subsubsection{Direct Selection for Modularity}
It has also been proposed that modularity may be directly selected for rather than indirectly as a side effect of stability or robustness. Rainey and Cooper considered the following situation \cite{Rainey2004}. An environment, referred to as environment 1, contains an unexploited niche which requires a major evolutionary innovation to be exploited. This innovation happens to have evolved in a different lineage in a different location, referred to as environment 2. Organisms from environment 2 may be transported into environment 1, in which they cannot survive. Thus, they lyse and their DNA is released into environment 1. In such a situation, cells in environment 1 with the greatest ability to accommodate this DNA will benefit the most since they can now exploit the new niche in environment 1. Thus, the ability to accommodate horizontally transferred genes confers a fitness advantage in the presence of unexploited niches and available
DNA from individuals. This ability will be reduced by pleiotropic effects and enhanced by modular genome architectures. Therefore, modularity will be directly selected for in the presence of horizontal gene transfer and ecological opportunity because it facilitates the accommodation of foreign DNA, which may be beneficial. The
authors emphasize that in their model modularity may increase evolvability but this is not the cause of the emergence of modularity.

\subsubsection{Templated Modularity}
A set of studies on the evolutionary emergence of modularity examined the evolutionary dynamics in a changing environment with goals that vary in a modular fashion such that each new goal shares subgoals with the previous goal. Such modularly varying goals have been found to lead to the spontaneous emergence of modularity.

In a first study \cite{Kashtan2005}, Kashtan and Alon studied the evolution of Boolean logic circuits and neural networks. When these systems were evolved under a fixed goal, they did not develop modularity. Even if the systems were started from a modular state, this modularity quickly decreased. On the other hand, if the systems were exposed to goals that periodically vary in a modular fashion, modularity did emerge and evolution proceeded faster. Randomly varying the environment did not lead to the emergence of modularity and resulted in a smaller increase in the rate of evolution than modularly varying goals.

An extension \cite{Kashtan2007} of this work analyzed the nature of the environmental variation in more detail by comparing modularly varying goals to different types of randomly varying goals. Here it was found that modularly varying goals generally lead a to large speedup of evolution while randomly varying goals may or may not increase the rate of evolution. The advantage of goals that vary in a modular fashion is more pronounced for more complex goals. As an explanation for this observation, Kashtan \emph{et al.}\ suggested that modularly varying goals can move populations away from local fitness maxima near which they may be stuck.

The results from the previous two paragraphs have been confirmed in an analytic model of evolution under modularly varying goals \cite{Kashtan2009a}. In this linear model, an evolving individual is represented by a matrix $\boldsymbol A$ which maps an input $\boldsymbol v$ to an output $\boldsymbol u$ by $\boldsymbol A \boldsymbol v = \boldsymbol u$. The fitness of the individual is defined as
\begin{align}
    F(\boldsymbol A) - F_0 = -\varepsilon\left\|\boldsymbol A\right\|^2 - \left\|\boldsymbol A \boldsymbol V - \boldsymbol U\right\|^2,
\end{align}
where the first term on the right-hand side ascribes a cost to the individual based on the magnitude of the elements of $\boldsymbol A$ and the second term represents a reward for correctly mapping a given set of inputs $\boldsymbol V$ to a given set of outputs $\boldsymbol U$. If $\boldsymbol A$ is block-diagonal or almost so, it will be considered modular. In this analytic model, Kashtan \emph{et al}.\ observed the same trends as before. Constant goals lead to non-modular structures and slow convergence, while modularly varying goals lead to modular solutions and fast convergence, especially for harder goals. If goals stop varying, modularity decreases, and randomly changing goals generally result in evolutionary confusion.

Most recently, Kashtan \emph{et al.}\ considered evolution in a spatially, rather than temporally, heterogeneous environment in the presence of extinctions \cite{Kashtan2009}. In this case they observed that without extinctions networks evolve to be highly optimal but not modular, whereas with extinctions modularity emerges. Here they suggest that modularity is selected for because it enables individuals to rapidly adapt to free niches after an extinction event. They also realized that, in contrast to their previous studies, recombination is very important for rapid adaption to new niches.

\subsubsection{Spontaneous Emergence of Modularity as a Phase Transition}
\label{sec:spontaneousEmergence}
In this section we will show
in general how modularity emerges spontaneously under a small set of assumptions. The argument is motivated by two observations. First, evolvability is a selectable trait and will be selected for in a changing environment \cite{Earl2004}. Second, modularity enhances the evolvability of organisms \cite{Bogarad1999a}. Consequently, the selection for evolvability leads to the emergence of modularity. We will first review an evolutionary model in which modularity emerges spontaneously.
In this model, a population of individuals evolves on a rugged fitness landscape,
the  individuals engage in horizontal gene transfer,
and the environment is changing in time.  We will then review some results that this model may explain.

Evolvability evolves \cite{Earl2004}. The evolvability of organisms is determined by the mutational processes acting on their genome and the rates at which these occur. Such processes may include point
mutation, recombination, transposition, or horizontal gene transfer. The capability to perform these processes as well as the rates at which they occur are encoded in the genome and, thus, under selective pressure. In a changing environment, organisms with adaptable evolutionary frameworks will have a fitness advantage over their peers that are not as adaptable. This advantage imposes a selection for adaptable frameworks and hence evolvability.

As described in the previous sections, modularity confers many benefits
to an evolving system. It allows for biological information to be stored in pieces or to be swapped in large chunks. It also enhances the robustness of a system and makes it more evolvable. However, do these benefits imply that modularity is inevitable? What is the probability for the emergence of modularity? Is modularity a typical or a special case? Here we explore the conjecture that modularity will spontaneously emerge in any evolving population which evolves on a rugged fitness landscape in a changing environment and undergoes horizontal gene transfer.

Lipson \emph{et al}.\ were among the first to describe a quantitative model for the emergence of modularity
\cite{Lipson2002}. 
 In a simple linear algebra model they suggested
that modularity arises spontaneously in response to variation.
However, Gardner and Zuidema analyzed the model of Lipson \emph{et al}.\ and came to the conclusion that it failed to establish a clear link between modularity and evolvability \cite{Gardner2003}. 
As we will see, one reason is that
Lipson \emph{et al}.\ did not consider horizontal gene transfer.

The spontaneous emergence of modularity has been observed in a generic evolutionary model described in \cite{Sun2007} and extended in \cite{He2009}. In this model  evolution is assumed to occur on a rugged fitness landscape with many local optima. Such a rugged landscape, which imposes a pressure for efficient evolutionary structures, can be generically described by a spin glass Hamiltonian \cite{He2009},
\begin{align}
    H^\alpha\left(s^{\alpha,l}\right) = \frac{1}{2\sqrt{N_D}} \sum_{i \ne j} \sigma_{i,j}\left(s^{\alpha,l}_i, s^{\alpha,l}_j\right) \Delta^{\alpha}_{i,j}.
    \label{eq:microscopicFitness}
\end{align}
Here $s^{\alpha,l}_i$ is a string by which ``individual'' $l$ can be identified. The index
$i$ is in the range $1 \le i \le N$, where $N$ is the length of the string. This is a generic model of evolution in which $s^{\alpha,l}_i$ can, for example, represent an amino acid in a protein sequence or a protein in the genome. The variable $l$ is a label for the different individuals in the population such that $1 \le l \le N_{\mathrm{size}}$, where $N_{\mathrm{size}}$ is the number of individuals. The interactions between the $s^{\alpha,l}_i$ are governed by a structure,or connection matrix $\Delta^{\alpha}_{i,j}$, whose possible forms are enumerated by $\alpha$ with $1 \le \alpha \le D_{\mathrm{size}}$. Here $D_{\mathrm{size}}$ is the number of possible structures. The connection matrix generically represents the structure of interactions such as, for example, protein folds, or regulatory constraints. The matrix $\sigma_{i,j}(s_i, s_j)$ is symmetric in $i$ and $j$ and represents the interaction strength. Its values are chosen from a standard normal distribution. These random couplings encode the effect of the environment. They can take on both negative and positive values, which corresponds to having both ferromagnetic and anti-ferromagnetic interactions, leading to frustration. As a consequence, the fitness landscape is rugged with a large number of local extrema. Choosing the $\sigma_{i,j}(s_i, s_j)$ in a correlated way can lead to a less rugged, or even smooth if all are positive and equal, landscape \cite{Sun2007}.
The structure $\alpha$ is described by the matrix $\Delta^{\alpha}_{i,j}$, a binary symmetric contact matrix. The number of connections in each structure and hence the number of non-zero elements in $\Delta^{\alpha}_{i,j}$ is constrained to be a fixed number $N_D$. This ensures that modularity cannot emerge as a consequence of an increasing number of connections. Connections can only be redistributed. 

Horizontal gene transfer (HGT) is restricted to predefined blocks of equal length and the rate of horizontal gene transfer and the mutation rate are approximately equal. The modularity $M$ is defined to be the number of non-zero elements in blocks along the diagonal of $\Delta^{\alpha}_{i,j}$. The size of the blocks is equal to the length of the HGT segments. Note that this modularity will have a non-zero value, $M_0$, even for random distributions of connections.

This model allows one to study not only the evolution of individuals in a variable environment, but also the evolution of structural connections $\Delta^{\alpha}_{i,j}$. A simulation thus models the evolution of a population of $D_{\mathrm{size}}$ structures, and for each structure a population of $N_{\mathrm{size}}$ sequences.

There are three different levels of evolutionary change in this model. First,
the sequences within each structure change most rapidly by point mutation and horizontal gene transfer. In each round, the 50\% of the population with the highest fitness within each structure are randomly duplicated, where the fitness of each individual is given by the spin glass Hamiltonian in equation \eqref{eq:microscopicFitness}. The value of the Hamiltonian is the energy, while the fitness is non-decreasing in the negative of the energy. 
Second, 
the environment changes
after $T_2$ rounds of mutation, recombination, and selection.
Environmental change
is accomplished by assigning a new value to each of the elements of $\sigma_{i,j}$ with probability $p$. Thus, $p$ represents the severity of environmental change and $1/T_2$ the frequency. 
Third, the evolution of the structures represents the slowest change. The
$\Delta^{\alpha}_{i,j}$ undergo mutation and selection every $T_3$ rounds. The fitness of the structures is obtained by averaging the fitness of the sequences in each structure over the $T_3/T_2$ environmental changes. Of the structures only the top 5\% of the population are selected for the next round, which are randomly amplified to maintain a constant population size and also mutated.

\begin{figure}[htbp]
    \begin{center}
        \includegraphics[width=0.8\textwidth]{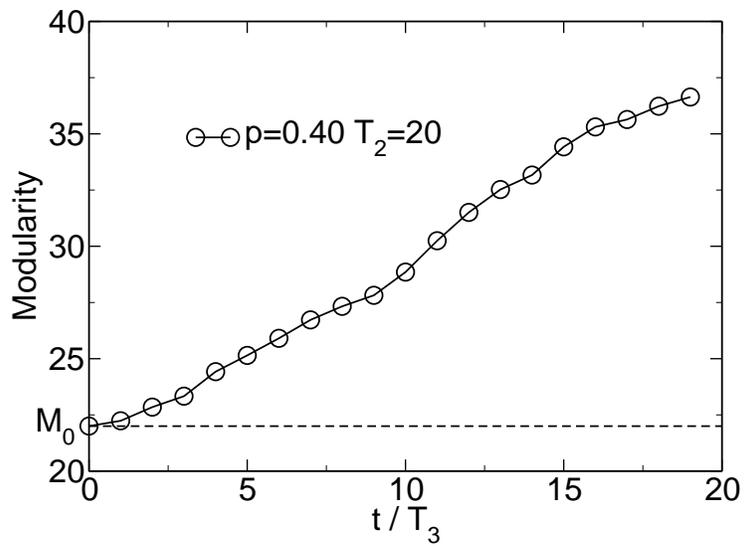}
    \end{center}
    \caption{Spontaneous emergence of modularity. $M_0 = 22$ is the baseline modularity for a random distribution of connections. From \cite{He2009}.}
    \label{fig:modIncWithTime}
\end{figure}

\begin{figure}[htbp]
    \begin{center}
        \includegraphics[width=0.8\textwidth]{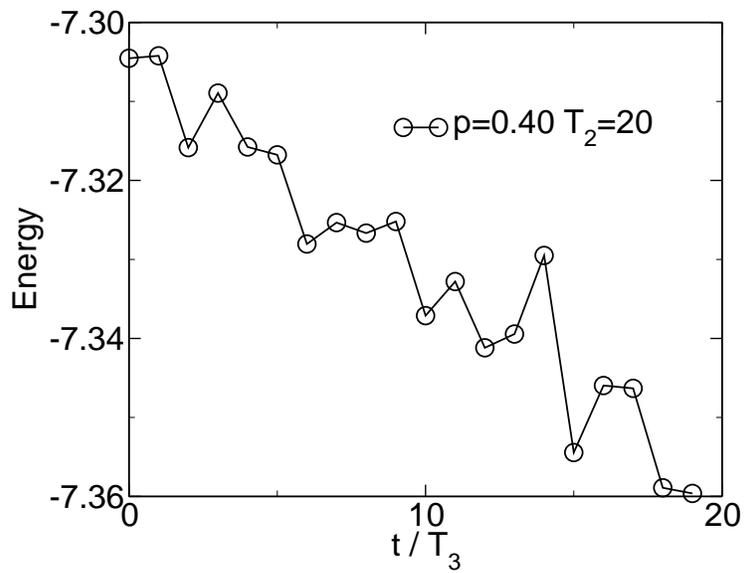}
    \end{center}
    \caption{The fitness (negative energy) increases with time and increasing modularity (see figure \ref{fig:modIncWithTime}). From \cite{He2009}.}
    \label{fig:fitIncWithTime}
\end{figure}

\begin{figure}[htbp]
    \begin{center}
        \includegraphics[width=0.8\textwidth]{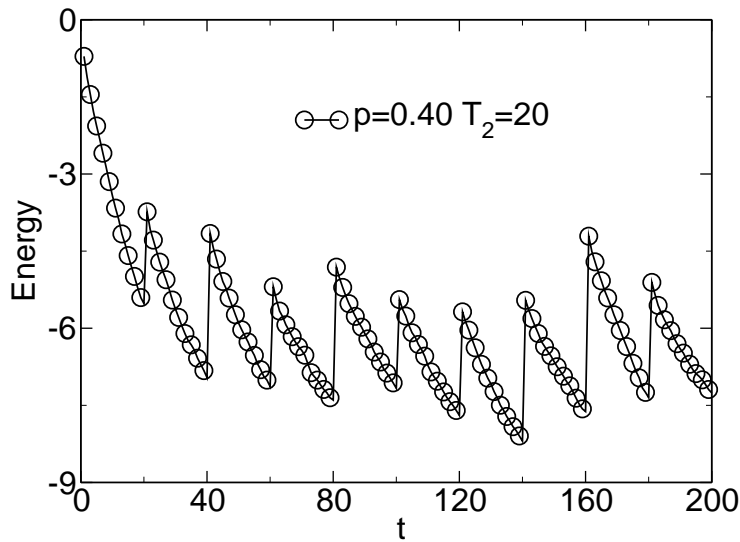}
    \end{center}
    \caption{Over short times, the fitness (negative energy) increases rapidly with time, but decreases significantly during environmental changes. From \cite{He2009}.}
    \label{fig:fitIncForShortTime}
\end{figure}

Simulations within this model show the emergence of modularity when
the environment is changing and when
horizontal gene transfer is present. Figure \ref{fig:modIncWithTime} shows the emergence of modularity for one instance of the model. $M_0 = 22$ is the modularity of the random state. The modularity grows linearly, because it is far from its steady-state value. This growth of modularity can be considered a symmetry-breaking event, where the order parameter is the excess modularity $M - M_0$. The symmetry being broken is the permutation symmetry of the connection matrix. 
It is the linear topology of the HGT event that allows this symmetry
to be broken.
The matrix moves from an initially uniform random distribution of entries to a modular distribution of entries clustered along the diagonal. Figure \ref{fig:fitIncWithTime} shows that over large time scales this increase in modularity is associated with an increase in fitness, measured as a decrease in energy. Figure \ref{fig:fitIncForShortTime} shows the change of fitness for short times. During times of constant environment, the fitness increases rapidly, but every time the environment changes, the fitness decreases substantially because the sequences are not well adapted to the new environment. These results are not sensitive to the values chosen for the parameters \cite{He2009}.

\begin{figure}[htbp]
    \begin{center}
        \includegraphics[width=0.8\textwidth]{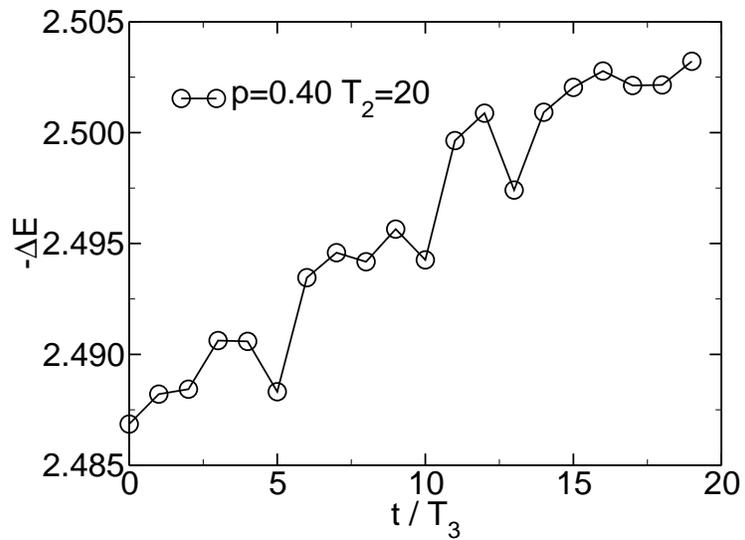}
    \end{center}
    \caption{The evolvability, as measured by the increase in fitness during evolution in one environment, increases with time as modularity increases. From \cite{He2009}.}
    \label{fig:evolvability}
\end{figure}

The evolution of evolvability can also be observed in this model. Evolvability can be measured by the increase in fitness while the sequences evolve in one environment. Over long time scales, the average gain in fitness between two subsequent environmental changes is not constant. As figure \ref{fig:evolvability} shows, there is a clear trend towards increasing evolvability over time. The evolved modular structure of the connection matrix allows the sequences to evolve faster in each new environment.

\begin{figure}[htbp]
    \begin{center}
        \includegraphics[width=0.8\textwidth]{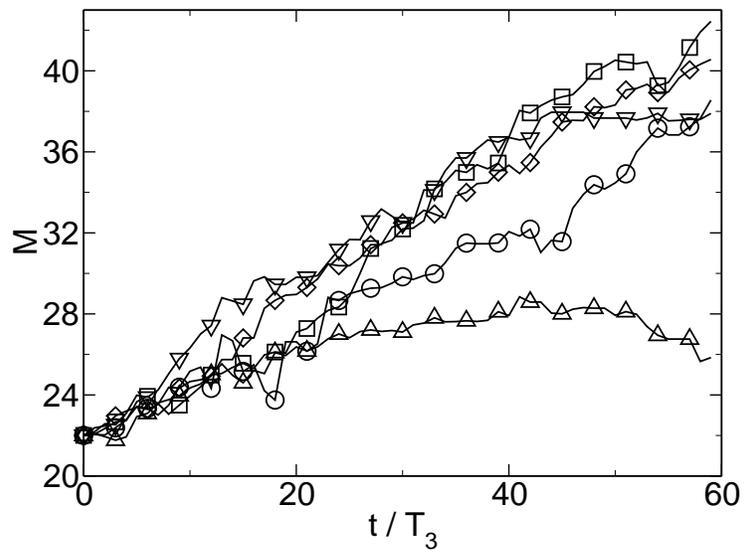}
    \end{center}
    \caption{Modularity growth for gene swaps of Poisson random length at a uniformly random position. The average swap lengths for the results shown are 10 ($\bigcirc$), 20 ($\Box$), 20 ($\Diamond$), 5 ($\bigtriangleup$), and 40 ($\bigtriangledown$). The number of attempted swaps are 12, 6, 12, 24, and 3, respectively. From \cite{He2009}.}
    \label{fig:randomSwap}
\end{figure}

These results robustly persist if a different initial contact matrix is used. Choosing a scale-free network yields almost identical results \cite{He2009}. Similarly, relaxing the biologically motivated constraint that horizontal gene transfer can only recombine predefined pieces of equal length does not hinder the emergence of modularity. For gene transfer that starts at any position with equal probability and swaps pieces of a Poisson random length, the evolution of modularity, shown in figure \ref{fig:randomSwap}, robustly shows the expected trend.

\begin{figure}[htbp]
    \begin{center}
        \includegraphics[width=0.8\textwidth]{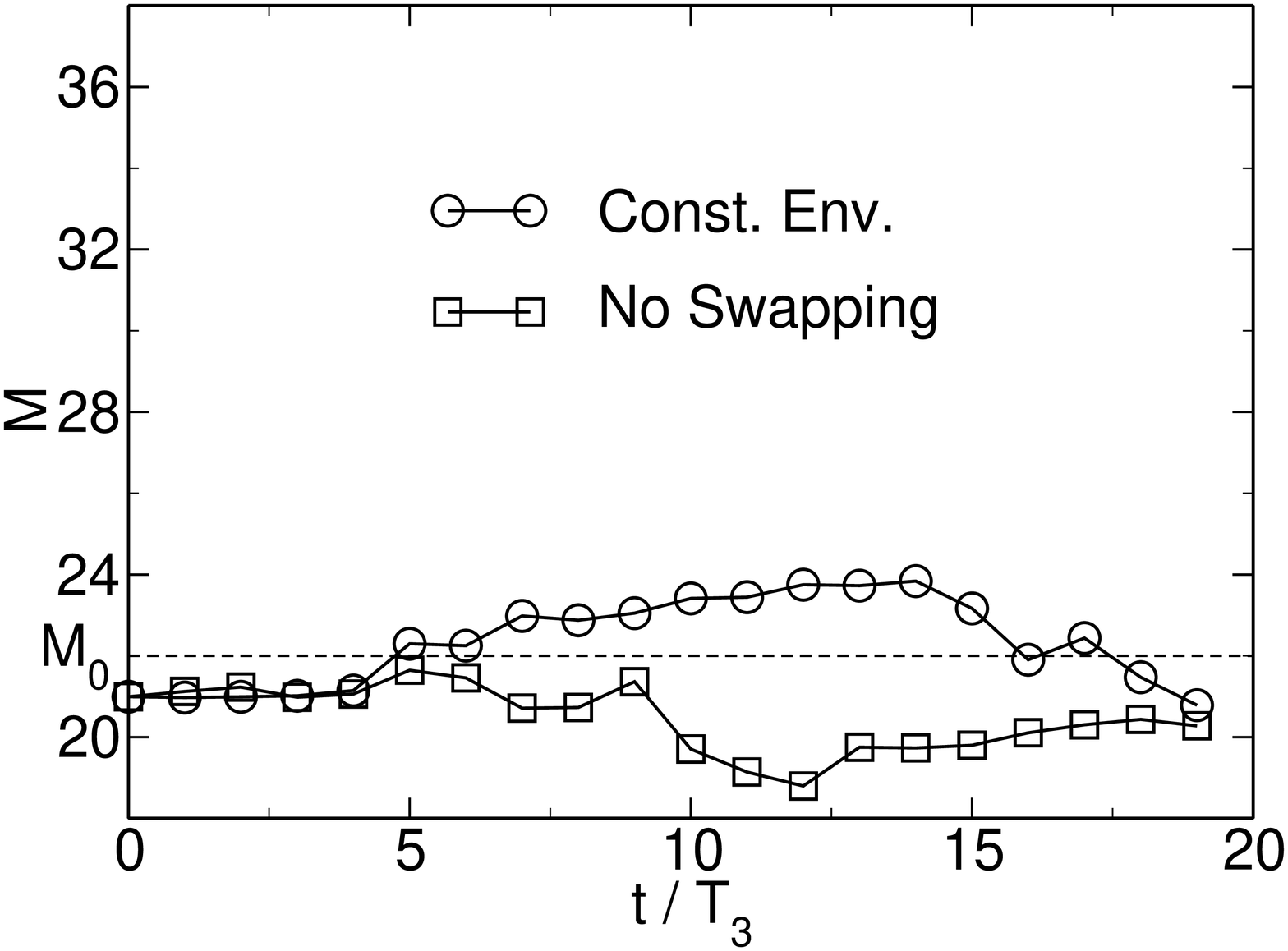}
    \end{center}
    \caption{If either environmental change or horizontal gene transfer is removed from the model, modularity does not emerge. From \cite{He2009}.}
    \label{fig:modConstEnvNoHGT}
\end{figure}

\begin{figure}[htbp]
    \begin{center}
        \includegraphics[width=0.8\textwidth]{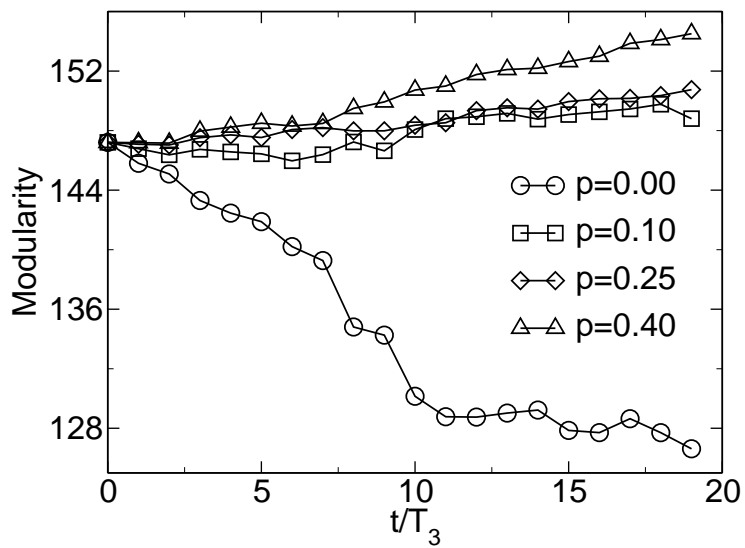}
    \end{center}
    \caption{Larger magnitudes of environmental change lead to a faster increase in modularity. From \cite{He2009}.}
    \label{fig:modVsAmp}
\end{figure}

\begin{figure}[htbp]
    \begin{center}
        \includegraphics[width=0.8\textwidth]{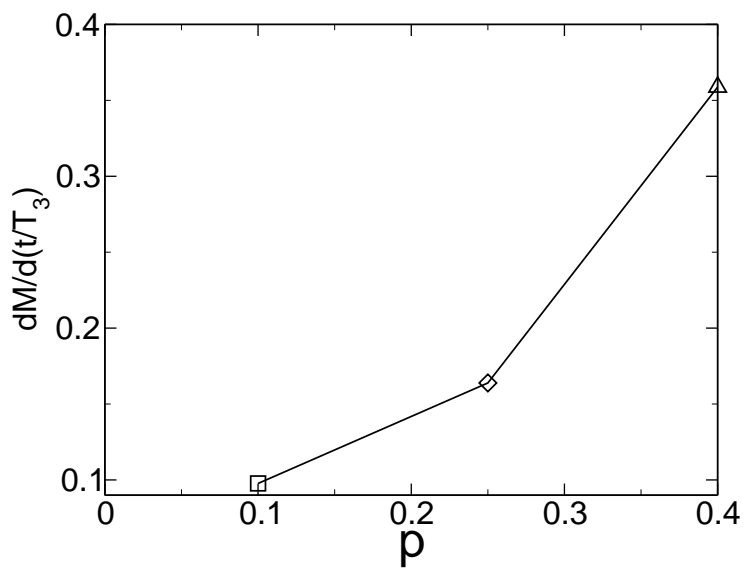}
    \end{center}
    \caption{The rate of change of modularity increases with increasing magnitude of environmental change. From \cite{He2009}.}
    \label{fig:dModVsAmp}
\end{figure}

Without environmental variability or without
horizontal gene transfer, however, no emergence of modularity is observed. As figure \ref{fig:modConstEnvNoHGT} shows, the modularity remains near $M_0$, the value for random networks. It is important to note that neither environmental change, nor horizontal gene transfer explicitly favor modularity. Rather, the system adopts a modular state under these conditions because modularity allows the system to respond better to the continuously changing environment. Thus, there is an implicit selection for evolvability in a variable environment, and horizontal gene transfer increases the evolvability of modular systems. In combination, horizontal gene transfer and environmental change implicitly select for modularity. One would therefore expect that the degree of modularity increases with increasing environmental change. Figure \ref{fig:modVsAmp} shows such a trend for varying degrees of the severity of environmental change. From an initially modular state, modularity decreases if there is no environmental pressure, while it increases in the presence of environmental change. For larger values of $p$, modularity increases more rapidly. This trend can be seen more clearly in the derivative of modularity with respect to time, shown in figure \ref{fig:dModVsAmp} for different severity of environmental change.

\begin{figure}[htbp]
    \begin{center}
        \includegraphics[width=0.8\textwidth]{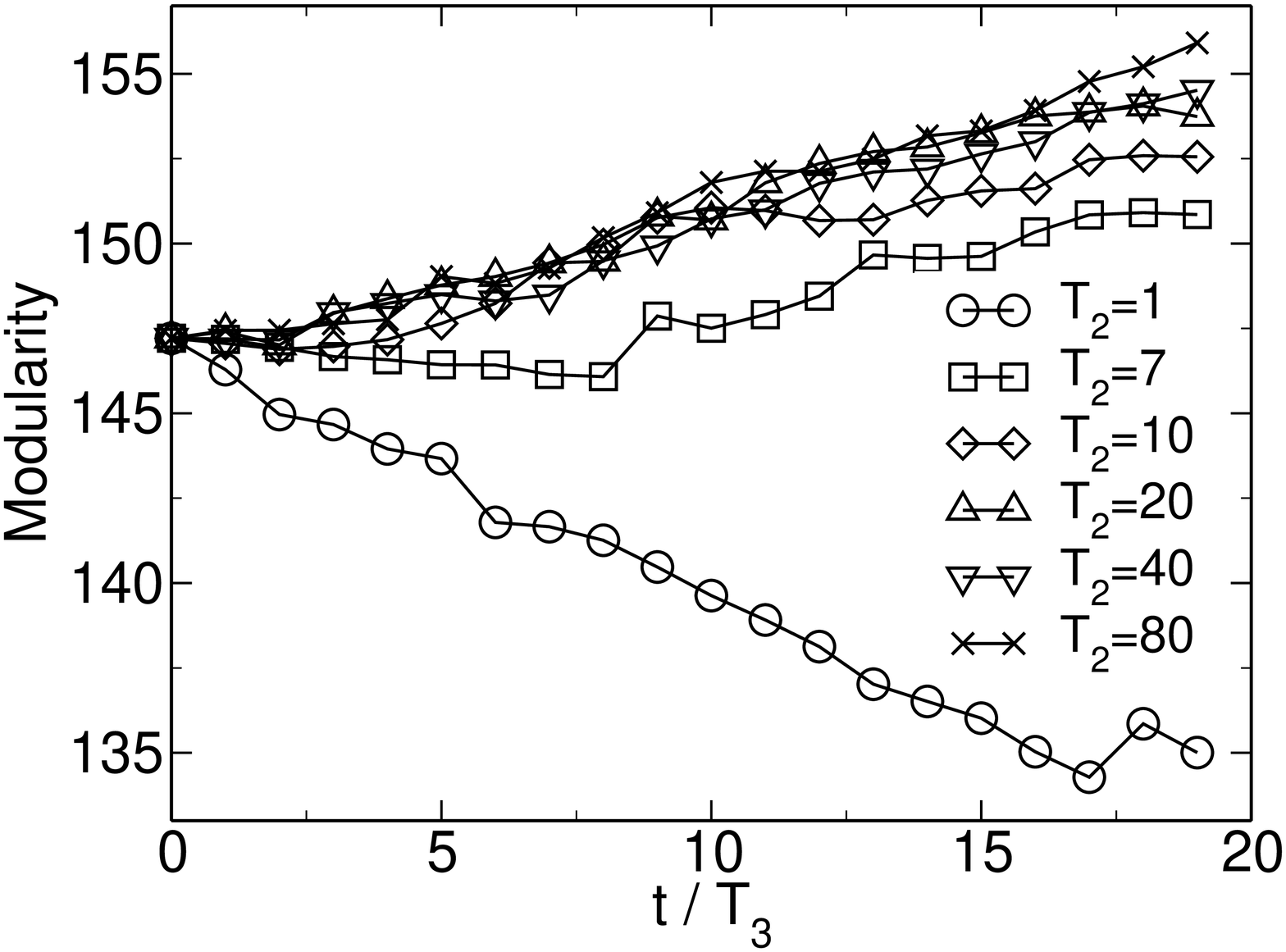}
    \end{center}
    \caption{More frequent environmental change leads to a faster increase in modularity. From \cite{He2009}.}
    \label{fig:modVsFreq}
\end{figure}

\begin{figure}[htbp]
    \begin{center}
        \includegraphics[width=0.8\textwidth]{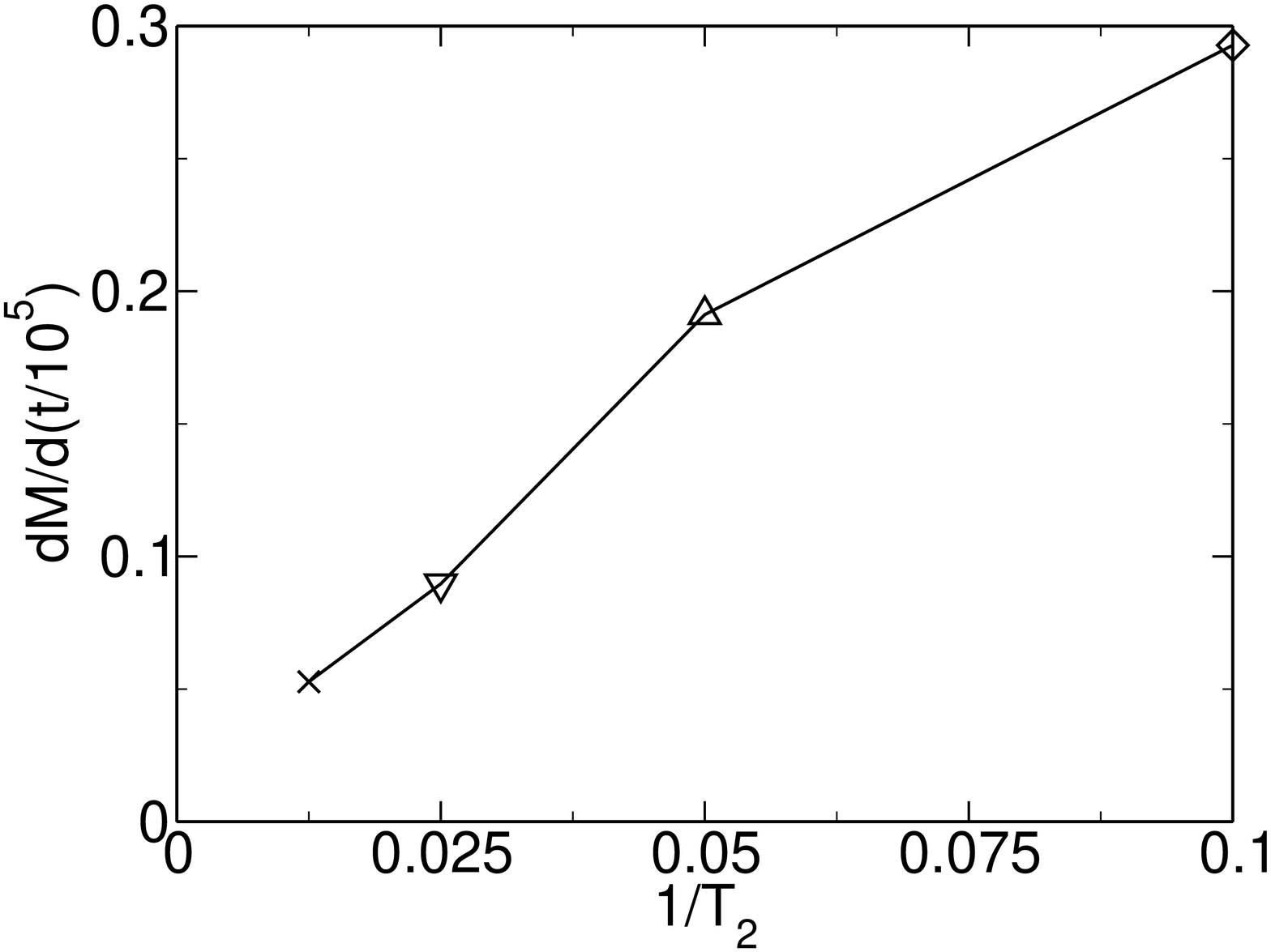}
    \end{center}
    \caption{The rate of change of modularity is approximately proportional to frequency of environmental change. From \cite{He2009}.}
    \label{fig:dModVsFreq}
\end{figure}

A similar observation can be made when varying the frequency of environmental change rather than the intensity. For very high frequencies, modularity decreases with frequency because the environment is changing too fast for the system to evolve in response to it. But, as figure \ref{fig:modVsFreq} shows, for moderate frequencies of environmental change, modularity increases with frequency just as it did with magnitude of environmental change. Figure \ref{fig:dModVsFreq} shows the rate of change of modularity versus frequency of environmental change.

\begin{figure}[htbp]
    \begin{center}
        \includegraphics[width=0.8\textwidth]{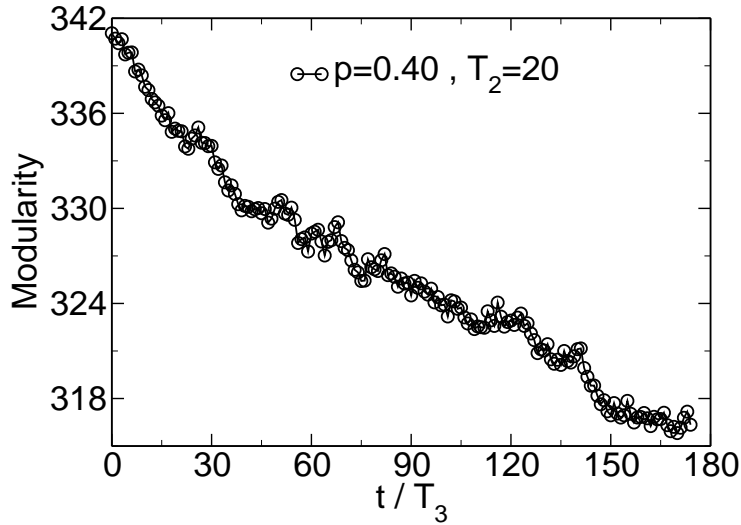}
    \end{center}
    \caption{If the modularity is initially above its steady-state value, it decreases with time because of the entropic effects of random mutations. From \cite{He2009}.}
    \label{fig:modDecay}
\end{figure}

The emergence of modularity is a response to the past variation in the environment of the system. Therefore, one can make the argument that in analogy to the fluctuation-dissipation theorem one would expect modularity to be proportional to the variance of the previously encountered environments \cite{He2009}. As figures \ref{fig:dModVsAmp} and \ref{fig:dModVsFreq} show, this can indeed be observed in this model. Similarly, in analogy to the competition between energy and entropy, one would expect there to be a steady-state value of modularity below the maximum modularity which depends on the parameters of the system. At this steady state there will be a balance between the entropic forces of random mutations driving modularity to its baseline value, $M_0$, and the selective forces which seek to enhance evolvability by increasing modularity. This effect can also be observed in the model by starting the system in a highly modular state. As figure \ref{fig:modDecay} shows, with time the modularity decreases from this very high value.

This model illustrates the spontaneous emergence of modularity in a population of evolving individuals under two conditions: the individuals can engage in horizontal gene transfer and the environment changes. This emergence of modularity is a symmetry-breaking event caused by the selection for evolvability in a changing environment \cite{Earl2004} and the fact that modular systems can take advantage of horizontal gene transfer to adapt to a new environment. The rate of modularity growth increases with the amplitude and frequency of environmental change. A constant environment does not promote the emergence of modularity. 

Other theoretical studies have been performed which support the results of this model. For example, Crombach and Hogeweg studied the evolution of simulated gene regulatory networks \cite{Crombach2008} and confirmed the observation that alternating environments lead to the evolution of evolvability. They found that even though mutations are random, their phenotypic effects become strongly biased by an evolving genotype-phenotype map. Martin and Wagner investigated the effects of recombination on models of transcriptional regulation circuits \cite{Martin2009}. Their findings include that the presence of recombination leads to the emergence of modular regulatory control, a reduction in the deleterious effects of mutations, and greater phenotypic diversity.

Misevic \emph{et al}.\ investigated how the reproductive mode shapes the genetic architecture of digital organisms \cite{Misevic2006}. In the theoretical model described in this section, modularity is expected to emerge in the presence of horizontal gene transfer, which enables the exchange of pieces of genetic information between different individuals. This leads to new genotypes that combine genetic material from two ``parent'' genotypes. Sexual reproduction also allows for large-scale exchange of genetic information and results in genotypes which are combinations of two parent genotypes. Hence, one would expect horizontal gene transfer and sexual reproduction to have similar effects. In their study Misevic \emph{et al}.\ found that sexual organisms have more modular and longer genomes, are more robust, and have a higher fitness \cite{Misevic2006}. In addition they observed that the strength of epistatic interactions is weaker in sexual organisms than in asexual ones and that the reproductive mode has a significant effect on the evolution of the genetic architecture. In a follow-up study \cite{Misevic2010}, they probed how changing environments influence the reproductive mode of these digital organisms. They report that in rapidly and strongly changing environments sexual reproduction becomes dominant regardless of the reproductive mode in which the population starts. Furthermore, in such environments predominantly sexual populations achieve a higher average fitness. Conversely, in slowly changing environments, asexual reproduction evolved to be predominant in most populations and sexual and asexual populations are equally fit on average.

Results very similar to the ones described in this section were obtained by Callahan \emph{et al}.\ in a later study who observed the spontaneous emergence of a type of modularity called collinearity in a computational model of the evolution of polyketide synthases \cite{Callahan2009}. This emergence was observed for a wide range of parameters, despite the fact that modularity provides no direct fitness benefit.  Modularity emerges only in the presence of continuous evolutionary pressure and horizontal gene transfer. This result was explained by a secondary selection effect because modularity increases the fitness benefits of recombination if the environment changes rapidly.

These studies provide evidence for the importance of environmental variation and horizontal gene transfer on the evolution of evolvability and modularity and are in agreement with the theory of spontaneous emergence of modularity
described in this section.

\section{Experimental Observation of Modularity}
In this section we review experimental observations in various biological networks that support the theory described in the previous section. We will present evidence for the evolution of evolvability, for the presence of modularity and the enhanced evolvability provided by modularity, and for the connection of environmental variation and horizontal gene transfer to the emergence of modularity.

\subsection{Modularity in Pathogens}
Pathogens are exposed to extreme environmental pressure and engage in extensive horizontal gene transfer. Therefore, we would expect them to evolve substantial modularity. Studies show that pathogens not only are very modular but also that this modularity enhances their evolvability by allowing them to vary mutation rates between different parts of their genome.

Structural and evolutionary modules have been observed in viruses. For example, Karlin \emph{et al}.\ found that \emph{Paramyxovirinae} are composed of six modules \cite{Karlin2003}. Viral proteins have also been shown to be modular. Ferron \emph{et al}.\ found modules by homology search in sequence data and ensured the validity of these modules by considering the results from other sources such as structure definition, biological data, and additional sequence properties
 \cite{Ferron2005}.
 The modular organization they discovered has helped to characterize virus domains by structure and function.
In influenza, it has been observed that the antibody immune response is
dominantly directed to the epitope regions of the 
hemagglutinin  protein on the surface of the virus particle.  While the mutation rate is
assumed to be constant throughout the RNA of the virus, the observed
substitution rate, or observed rate of evolution, in these five epitope
regions is significantly greater \cite{Gupta2006}.

All organisms balance the need for stability against the need of variability. Low rates of genetic change decrease deleterious mutations, but they also reduce the probability of beneficial mutations that may be necessary for adaptation. The optimal rates of genetic moves may vary with time or space. Radman \emph{et al}.\ observed that bacteria and viruses have evolved the ability to respond to these variations by genetically controlling their mutation and recombination rates to adapt to changing environments \cite{Radman1999}. The bacterial SOS response provides a prototypical example: under genotoxic or metabolic stress, bacteria will start to express mutator genes and upregulate several recombination genes \cite{Radman1999}. This evidence that rates of genetic change are under selective control supports the idea that evolvability can evolve.

The hypermutation described in the previous paragraph can also be limited to select parts of the genome, making it particularly useful for pathogens, which can increase their mutation and recombination rates at sites encoding surface antigens to escape the host's immune system. This result has been confirmed by other studies reviewed by Massey and Buckling who proposed that the constantly changing environment of pathogens selects for mechanisms which can generate phenotypic variation \cite{Massey2002}. A generalized hypermutation would lead to a significantly greater mutational load than the localized hypermutation of only those parts of the genome involved in interactions with the host environment \cite{Moxon2006}. Allocating mutation and recombination rates in a modular fashion across the genome enhances evolvability under excessive environmental pressure.

The positive effect of environmental variation on evolvability was also observed by Kepler and Perelson who showed in a differential equation model of virus dynamics that the presence of compartments with different drug concentrations increases the likelihood that a resistant strain emerges \cite{Kepler1998}. Especially for high drug concentrations, resistance was found to
emerge with a higher probability in the presence of spatial heterogeneity. Thus, environmental variability improves the evolvability of viruses.

\subsection{Modularity in Metabolic Networks, Gene Networks, and Protein-Protein Interaction Networks}
Since the seminal paper by Hartwell \emph{et al}.\ \cite{Hartwell1999}, the concept of modularity has been firmly established in cell biology and with it the idea that modular structures may facilitate evolutionary change. Advances in genomics and proteomics have allowed the creation of large data sets from which networks can be constructed. Many of these networks have been shown to posses a hierarchical modular structure. Care needs to be taken when analyzing networks obtained from databases of biological interactions, such as protein-protein interactions. It is difficult to estimate the error rate of experimental techniques and sometimes the overlap of the results between interaction data obtained through two different methods can be small \cite{Shoemaker2007}. Promising attempts have been made to judge and improve the quality of protein-protein interaction data by combining the results from different methods \cite{Braun2009}. This approach could also enhance other experimentally determined biological networks. In this section we review some of the evidence for the presence of modularity in metabolic networks, gene networks, and protein-protein interaction networks and the relationship between
modularity and environmental variation and horizontal gene transfer.

\subsubsection{Metabolic Networks}
Ravasz \emph{et al}.\ were among the first to investigate the structure of metabolic networks in detail
\cite{Ravasz2002}.
They found that metabolic networks have a scale-free architecture and are nevertheless highly clustered.
 Furthermore, the clustering coefficient in metabolic networks is independent of size, which is in stark contrast to random scale-free networks in which the clustering coefficient decreases with size. To explain these surprising findings, Ravasz \emph{et al}.\ suggested a new algorithm that can generate networks with topological properties in agreement with empirical metabolic networks. The resulting networks have a hierarchical modular structure \cite{Ravasz2002}. In 2006, Spirin \emph{et al}.\ extended the study of metabolic networks to include evolutionary information from genomic data that allowed them to analyze both functional and evolutionary modules \cite{Spirin2006}. In this metabolic-genomic network they also found modules on different scales, indicating hierarchical modularity. DaSilva \emph{et al}.\ introduced a parameter called ``core coefficient'' to quantify hierarchical modularity in networks and found that the core coefficient in metabolic networks significantly exceeds that of random networks \cite{daSilva2008}.

\begin{figure}[htbp]
    \begin{center}
        \includegraphics[scale = 0.8]{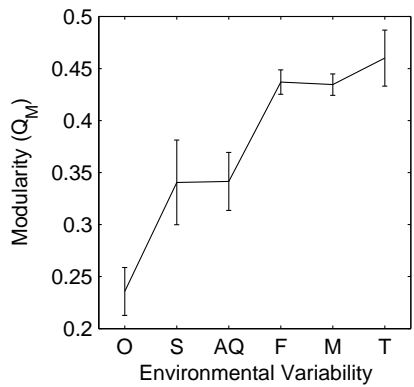}
    \end{center}
    \caption{Following \cite{Parter2007}: Modularity increases with environmental variability. Bacteria are grouped by lifestyle and ordered from lowest to highest environmental variability. O stands for obligate, S for specialized, AQ for aquatic, F for facultative, M for multiple environments, and T for terrestrial.}
    \label{fig:modVsEnvVar}
\end{figure}

After it was recognized that metabolic networks 
exhibit modularity, the question arose whether  metabolic networks are more modular in some organisms than in others. Parter \emph{et al}.\  probed this question by analyzing the relationship between environmental variability and modularity in the metabolic networks of more than one hundred species of bacteria \cite{Parter2007}. Their study showed that there is a positive correlation between environmental variation and modularity, shown in figure \ref{fig:modVsEnvVar}.
This correlation remains significant even after the difference in network size is taken into account. In addition, modularity was shown to have a stronger correlation with environmental variability than with phylogenetic proximity \cite{Parter2007}. Both of these results support the hypothesis that metabolic networks of organisms under greater environmental pressure evolve to be more modular.

Not only the importance of environmental change, but also the relation between horizontal gene transfer and modularity has been supported by empirical findings in metabolic networks. In 2008, a study by Kreimer \emph{et al.}\ explored modularity in more than three hundred bacterial metabolic networks and found three main determinants of modularity: network size, the environment, and horizontal gene transfer \cite{Kreimer2008}. The extent of horizontal gene transfer was obtained from \cite{Nakamura2004}. There it was measured by computing the probability that a DNA segment is extrinsic using Bayesian inference. A gene segment was considered extrinsic to a recipient if its nucleotide composition was significantly different from the rest of the recipient's genome while matching the nucleotide composition of a donor. 


Recently it has been pointed out that the dependence on network size may be a an artifact of the method used to compute modularity \cite{Good2010}. It was argued that modularity as conventionally defined will tend to increase with an increasing number of modules or nodes. This is a consequence of the null model implicit in the definition of modularity: a random graph that has the same degree sequence as the network under investigation. Since the probability that an edge will fall within a given module in this random network decreases with increasing network size, a larger network will tend to get a higher modularity score \cite{Good2010}. This can also be understood by realizing that stochastic noise will allow an algorithm to detect modules in any network, especially in sparse networks. For larger networks, there will be more 
such noise-induced modules, giving the impression of greater modularity.
\begin{figure}[htbp]
    \begin{center}
        \includegraphics[scale = 0.3]{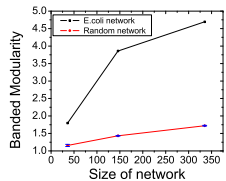}
    \end{center}
    \caption{The modularity of random networks increases with size. From \cite{He2009}.}
    \label{fig:heModRandom}
\end{figure}
See figure \ref{fig:heModRandom} for an example of this size-dependent modularity in random networks. From this figure is clear, however, that the modularity
observed in the natural network is significantly greater than 
the noise-induced value observed in the background model of a random network. Thus, any comparison of modularity in networks of different size or number of edges always needs to consider the values of modularity one would expect to obtain in a distribution of random networks with the same size and sparsity. However, the correlations observed between modularity and horizontal gene transfer and between modularity and  environmental effects should not be affected by this methodological bias. In particular, the data show that even bacteria with small metabolic networks, in which modularity may be underestimated by the algorithm, exhibit high modularity if they live in highly variable environments \cite{Kreimer2008}. It was also shown that bacteria occupying a limited number of niches have less modular metabolic networks than species occupying a greater variety of niches and pathogens that alternate between hosts have more modular metabolic networks than do single-host pathogens \cite{Kreimer2008}. Thus, environmental variability and horizontal gene transfer seem to be closely related to the modular structure of bacterial metabolic networks.

\subsubsection{Gene Networks}
In gene networks, the advantages of a modular architecture become highly apparent. The possibility of using novel combinations of modules rather than evolving new genes from scratch greatly enhances evolvability. A genomic study has shown that such a modular rewiring contributed significantly to the evolution of new functionality, especially in the evolution of proteins
 \cite{Patthy2003}.
Moreover, it was reported that the loss of intermodule introns inhibits further modular evolution.

Segr\`e \emph{et al.}\ constructed an epistatic interaction network from the results of all single and double knockouts of almost one thousand metabolic genes in \emph{S. cerevisiae}
 \cite{Segre2005}. 
 They found that this network consists of modules of genes arranged in a hierarchy, where modules are clusters of genes which interact monochromatically (either all aggravating or all buffering).
Based on this observation they suggested to extend the concept of epistasis from genes to functional modules, which interact epistatically as a group.
In other words, a second-order modularity of epistasis had emerged.

Bhattacharyya \emph{et al}.\ reviewed the importance of modular interactions in cell signaling circuits
 \cite{Bhattacharyya2006}.
They proposed that in such circuits, modularity may contribute to evolvability by making the evolution of new complex circuits and resulting phenotypes easier. This availability of new phenotypes would be especially beneficial in competitive and changing environments and may explain how modularity is maintained despite nonmodular systems often having a higher fitness in an unchanging environment.
 Modularity in gene networks may also contribute to an organism's robustness and the ability to maintain homeostasis \cite{Wang2008}.

Following the theory developed in section \ref{sec:spontaneousEmergence}, we would expect modularity to emerge in organisms in response to environmental pressure. Similarly, we would anticipate that systems which are directly involved in interactions with the environment would evolve to be more modular than systems which have no external interactions. This prediction has been verified by Singh \emph{et al.}\ in an evolutionary study of three bacterial stress response networks \cite{Singh2008}. They observed that the regulatory network for chemotaxis, a process which allows an immediate response to the environment, has greater modularity than that for sporulation, which is more indirectly affected by the environment. Furthermore, the network regulating DNA uptake, which is hardly impacted by the environment, displays no significant modularity. These results illustrate the influence of environmental change on the emergence of modularity in stress response networks.

Gene regulation can be considered a higher order modularity, and it was mentioned at the end of the previous subsection on metabolic networks. Modularity in gene regulatory networks decreases the complexity of the circuitry required for complex responses to external stimuli \cite{McAdams2004}. Transcriptional regulation factors are often acquired through horizontal gene transfer \cite{Price2008} and it has been observed that regulatory circuits evolves faster than the genes they regulate \cite{Lozada-Chavez2006, Maslov2004}. Thus, the higher order modularity displayed by gene regulation enhances evolvability.

\subsubsection{Protein-Protein Interaction Networks}
\begin{figure}[htbp]
    \begin{center}
        \includegraphics[width = 0.6\textwidth]{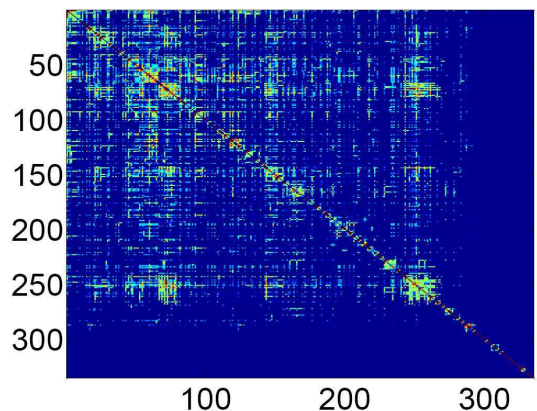}
    \end{center}
    \caption{Topological overlap matrix of the \emph{E.\ coli} protein interaction network. From \cite{He2009}.}
    \label{fig:ecoli_122_2}
\end{figure}

Modularity has also been observed in protein-protein interaction networks. An example of a clustered protein-protein interaction network with visually discernible modules is shown in figure \ref{fig:ecoli_122_2}. One of the first studies of protein-protein interaction networks on a meso-scale level was carried out by Spirin \emph{et al}.\ in 2003 who discovered highly statistically significant modules \cite{Spirin2003}. Functional modules in protein interaction networks can also be found from sequence data alone and agree with modules found by other methods
 \cite{vonMering2003}. 
 This indicates that the modularity in protein networks is encoded in the genome.
A study that incorporated data from a variety of sources including gene expression, functional annotations, evolutionary conservation, and protein structure supports the observation of a modular topology in protein interaction networks \cite{Gavin2006}.

Han \emph{et al}.\ studied the modular structure of protein networks in more detail by considering their temporal changes
 \cite{Han2004}.
 They found two types of hubs which they named ``party hubs'' and ``date hubs.'' While party hubs interact with most of their partners at the same time, date hubs bind their partners at different times or places.
 They investigated the topological role these different types of hubs play in the protein network. They observed that party hubs function mainly inside of modules while date hubs act as global connectors between modules. Consequently, the network is less resilient to the removal of date hubs than it is to the removal of party hubs. This hierarchy of hubs provides evidence for hierarchical modularity in protein-protein interaction networks.

The influence of the environment on modularity that was seen in metabolic networks and gene networks has also been noted in protein-protein interaction networks. A study by Cohen-Gihon \emph{et al}.\ showed that a protein with a function that is common to all organisms exhibits a lower degree of structural modularity than a protein that can only be found in few cell types \cite{Cohen-Gihon2005}. Campillos \emph{et al}.\ obtained even more explicit evidence for the effects of changing environments and horizontal gene transfer on modularity by studying evolutionarily cohesive functional modules in protein networks \cite{Campillos2006}. This allowed them to compare modules by evolutionary age. They found that young modules are frequently horizontally transferred between species.
These young modules are enriched in functions related to interactions with the environment.  These young modules also play an important role in the adaptation to new environments of species. Ancient modules, on the other hand, are often very well conserved and enriched in core functions such as metabolism and information processing. Furthermore, bacteria living in competitive, varying, and stressful environments acquired the most modules \cite{Campillos2006}. These observations clearly demonstrate how important horizontal gene transfer and environmental heterogeneity in space or time are for the presence of modularity.

A recent piece of evidence for the hypothesis that modularity emerges spontaneously in the presence of environmental change and horizontal gene transfer comes from a quantitative study of the evolution of modularity across evolutionary time scales \cite{He2009}. In this paper, a measure of protein divergence time was used to show that modularity in protein interaction networks has increased with time. The evolutionary age of proteins was quantified using the concept of compositional age. This method considers proteins to be older if they contain a larger fraction of older amino acids. The calibration of this measure using known divergence times enabled the construction of a mapping between compositional age and real age. To quantify modularity, topological overlap matrices were constructed from the interaction networks and reordered using average linkage hierarchical clustering. Modularity was computed using several different quantitative definitions and normalized by network size. For all definitions it was robustly observed that modularity has grown throughout evolutionary time in both organisms that were studied \cite{He2009}. These results are consistent with the theory that environmental change and horizontal gene transfer naturally lead to an evolution of increased modularity.

\begin{figure}[htbp]
    \begin{center}
        \hspace{0.315\textwidth}\includegraphics[width = 0.6667\textwidth]{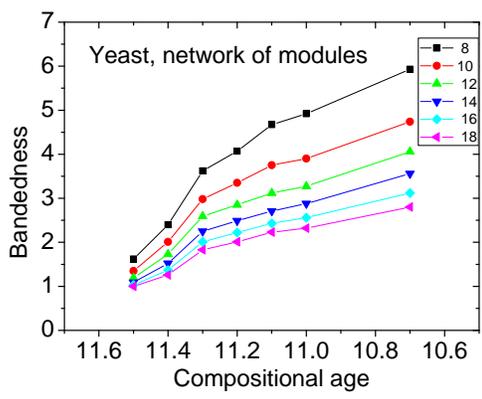}  \\
        \includegraphics[width = \textwidth]{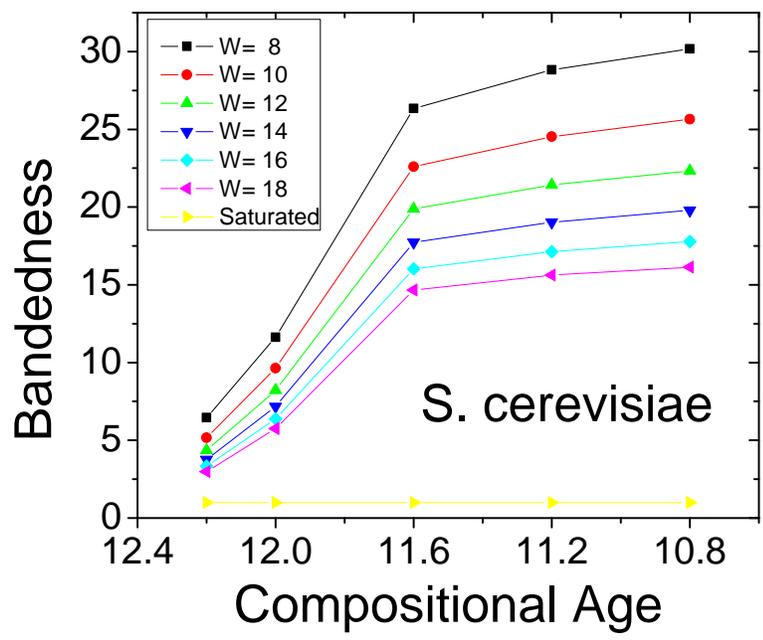}
    \end{center}
    \caption{Upper graph: second order modularity, lower graph: first order modularity from \cite{He2009}. The second order modularity continues to increase after the first order modularity saturates. The compositional age axes have been aligned.}
    \label{fig:heModularity}
\end{figure}

In reference \cite{He2009}, the graphs displaying the growth of modularity as a function of evolutionary time show that modularity, after increasing initially, seems to saturate. Considering that the rate of evolution is believed
to increase rather than saturate, this is at first a
 surprising result. To understand this observation, we extended the analysis presented in \cite{He2009} by computing a second order modularity. We constructed a new weighted network whose nodes are the projections of
the modules of the original network, that is we carried out one-step of
 a renormalization group operation. If, in the original network, nodes in one module had connections to nodes in other modules, then the vertices corresponding to these two modules in the new network were connected. The edges in the new network were assigned weights equal to the number of such inter-module connections in the original network. We then quantified the modularity in this new network of modules by the same method described in \cite{He2009}. We observed that this second order modularity also increases with evolutionary time, as shown in figure \ref{fig:heModularity}. Notably, it continues to rise even after the first order modularity seems to have saturated. This may indicate the emergence of a higher level of structure.

All the findings described above for metabolic networks, gene networks, and protein-protein interaction networks are in agreement with and can be explained by the theory of spontaneous emergence of modularity
described in section \ref{sec:spontaneousEmergence}. In the presence of horizontal gene transfer and environmental pressure, modularity will emerge spontaneously in a population of evolving individuals.

\subsection{Modularity in Ecological Networks}
\label{sec:modularityInEcology}
Ecological networks summarize interactions among all species in an ecosystem, represented as nodes, and the biotic interactions between them, represented as edges between the nodes. There are typically two types of interactions that can exist between species: mutualistic or trophic. A trophic link between two species indicates that one eats the other, an antagonistic relationship, while a mutualistic link indicates a relationship in which both species benefit, such as between plant and their pollinators or seed dispersers. Most empirical studies of ecological networks focus on only one of these types of interaction which results in either trophic networks (food webs) or mutualistic networks (\emph{e.g.}\ pollination networks).

Topological analysis has revealed that most food webs have a hierarchical structure and some can be decomposed into compartments or modules \cite{Krause2003}. This modularity increases the stability of food webs \cite{Thebault2010} by localizing the impact of a disturbance within a single compartment and minimizing impact on other compartments \cite{Krause2003}.

We first describe why mutualistic networks are not expected to be modular. A recent study by Thebault and Fontaine found that the relation between network architecture and stability is fundamentally different between mutualistic networks and trophic networks 
\cite{Thebault2010}. 
While in the latter compartmentalization increases stability, it has the opposite effect in the former. The difference 
between trophic and mutalistic networks
is the difference between a rugged and a smooth fitness landscape. Trophic interactions lead to frustration while mutualistic interactions do not. One model of evolution of ecological networks is the ``tangled nature'' model introduced in \cite{Christensen2002}.In this model, the evolution is governed by a replication rate, or microscopic fitness, which is very similar to the negative of the spin-glass Hamiltonian \eqref{eq:microscopicFitness}:
\begin{align}
    H(\boldmath S^\alpha, t) = \frac{1}{N(t)} \sum_{\boldmath S \in \mathscr S} n(\boldmath S, t) \sum_{i=1}^L J_i (\boldmath S^\alpha, \boldmath S) S_i^\alpha S_i
        - \mu N(t).
\end{align}
Here $t$ is the time, and the vector $S^\alpha$, whose elements can only take on the values $\pm1$, describes an individual. The sum over $\boldmath S$ runs over the entire genome space $\mathscr S$, $N(t)$ is the population size at time $t$, and $n(\boldmath S, t)$ is the occupancy of position $\boldmath S$ at time $t$ (how many individuals have genotype $\boldmath S$ at time $t$). The interaction matrix $\boldmath J^{ab} = \boldmath J(\boldmath S^a, \boldmath S^b)$ does not change with time and describes the interactions (trophic, mutualistic, or competitive) between an individual with genotype $a$ and and individual with genotype $b$. At any given time only a fraction of the entire genome space is occupied and hence contributes to $H$. The second term $\mu N(t)$ describes the limited availability of resources in the environment, where $\mu$ is the mean sustainable population size. The main differences between the tangled nature model and the spin-glass Hamiltonian \eqref{eq:microscopicFitness} are that in the former the interactions are not symmetric --- an interaction between two individuals can benefit one individual, but harm the other --- and that the fitness of each individual depends on the occupation number of all positions in genotype space to which it is connected through a non-zero interaction term $J_i$.

In a mutualistic network, all interactions are beneficial for both parties, and hence, the values of all $J_i$ are positive. This is analogous to having only ferromagnetic couplings, in which case there is no frustration and it is easy to find the ground state. In a trophic network, however, interactions are always beneficial for one party and detrimental for the other. This is comparable to a spin-glass with ferromagnetic and anti-ferromagnetic couplings, which is characterized by frustration and slow dynamics. This difference in the dynamics between mutualistic and trophic networks has been observed numerically in variations of the tangled nature model. Rikvold and Sevim studied \cite{Rikvold2007,Rikvold2007a} the distribution of the durations of quasi-steady states in mutualistic and trophic networks obtained from simulations and found that both follow a power law. The power-law exponent for the mutualistic case is more negative than that of the trophic case, which indicates that the mutualistic network is characterized by faster dynamics, as it would be expected for evolution on a smooth landscape. It was also observed that trophic interactions lead to hierarchically structured networks in the simulations while mutualistic interactions do not \cite{Rikvold2007,Rikvold2007a}. As discussed in section \ref{sec:spontaneousEmergence}, if evolution occurs slowly on a rugged fitness landscape, we expect the emergence of modularity. However, if evolution occurs rapidly, as it does on a smooth landscape, the discussion at the beginning of section \ref{sec:smoothLandscape} suggests that we should not expect emergence of modularity. By this independent line of reasoning, the results of \cite{Rikvold2007,Rikvold2007a} also suggest that the dynamics in mutalistic networks are rapid.
Thus, modularity is not expected to spontaneously arise in mutalistic
networks.  

On the other hand, based on the general theory proposed in section \ref{sec:spontaneousEmergence}, we expect that food webs under greater environmental pressure might evolve to become more hierarchical. To test this hypothesis we investigated hierarchy in 22 empirical food webs from rivers in New Zealand. The data was assembled by Thompson and Townsend (\emph{e.g.}\ \cite{Thompson2003, Thompson2005}) and provided on the Interaction Web Database%
\footnote{\texttt{http://www.nceas.ucsb.edu/interactionweb/}}.
We restricted our study to a limited geographical region to reduce the effect of potential confounding factors on food web architecture such as latitude or biome. Including only river food webs also minimizes the influence different habitats may have on network topology. As a proxy for environmental pressure we considered the availability of energy from detritus (particulate organic matter) input. Detritus is central to many food webs \cite{Sabo2005} and particularly small rivers are fueled by detritus input from surrounding terrestrial plants \cite{Vannote1980}. Thompson and Townsend showed that the forest river food webs in this data set have a much greater concentration of both coarse and fine particulate organic matter than their counterparts flowing through grassland \cite{Thompson2005}.

Our analysis proceeded as follows. We used the Euclidean commute time as a distance metric, defined for each pair of nodes as the expected time it takes a random walk to travel from one of the nodes to the 
other and back \cite{Luxburg2007}. 
This Euclidean commute time between the nodes of a weighted graph decreases when the number of paths connecting two nodes increases.
The commute time between two nodes also decreases when the length of any path connecting the
nodes decreases.
These properties make the Euclidean commute time well-suited for clustering tasks \cite{Saerens2004}. Let $L$ denote the graph Laplacian, defined as $L = D - A$, where $A$ is the adjacency matrix and $D = \operatorname{diag}(A_i)$ with $A_i = \sum_j A_{ij}$ is the degree matrix, a diagonal matrix whose elements are the degrees of the nodes. It is shown in \cite{Saerens2004} that the computation of the average commute time can be obtained from $L^+$, the Moore-Penrose pseudoinverse \cite{Barnett1990} of the graph Laplacian $L$, by
\begin{align}
    n(i,j) = V_G \left(\boldsymbol{e}_i - \boldsymbol{e}_j\right)^T L^+ \left(\boldsymbol{e}_i - \boldsymbol{e}_j\right).
\end{align}
Here $(\boldsymbol{e}_i)_j = \delta_{ij}$ and $V_G = \sum_{ij} a_{ij}$. Since it can be shown \cite{Saerens2004} that $L^+$ is symmetric and positive semidefinite,  $T_{ij} = [n(i,j)]^{1/2}$ is a Euclidean distance metric, called the Euclidean commute time (ECT) distance.

\begin{figure}[htbp]
    \begin{center}
        \includegraphics[scale = 0.8]{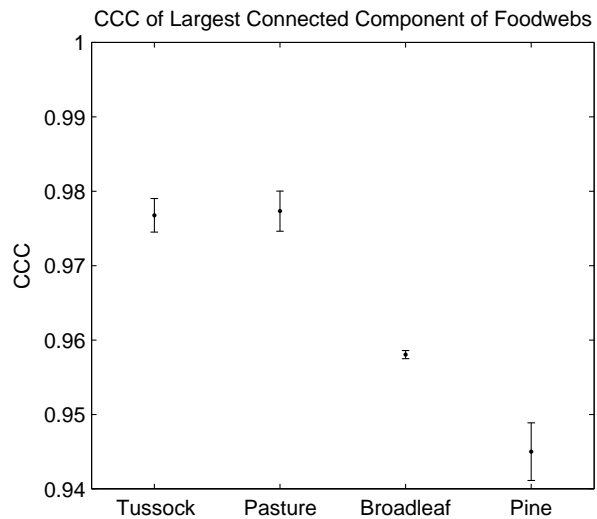}
    \end{center}
    \caption{Cophenetic correlation coefficient for river food webs surrounded by different terrestrial habitats. Food webs of rivers flowing through habitats with low detritus input are more hierarchical. 
For each habitat, the error bars are one standard error.
Data from \cite{Thompson2003, Thompson2005}. 
}
    \label{fig:cccFoodWebs}
\end{figure}

After finding the commute distances, we performed average linkage
hierarchical clustering on the matrices of commute distances to build a
hierarchy of clusters. Finally, we quantified hierarchy by computing the
cophenetic correlation coefficient (CCC) for each network. The CCC is a
measure of how well the dendrogram distances correlate with the original
commute distances. The CCC is defined as the Pearson correlation coefficient
between the node-node distances in the original data and that
in the tree-like representation:
\begin{align}
   \operatorname{CCC} = \frac{\displaystyle \sum_{i < j} \left(T_{ij} -
T\right)\left(c_{ij} - c\right)}
           {\displaystyle \sqrt{\sum_{i < j} \left(T_{ij} - T\right)^2
\sum_{i < j} \left(c_{ij} - c\right)^2}},
\end{align}
where $T$ is the average of the commute distances, $T_{ij}$, and $c$ is the
average of the dendrogram distances, $c_{ij}$. Our results are shown in
figure \ref{fig:cccFoodWebs}. The CCC values are significantly different
between rivers surrounded by forest (pine, broadleaf) and rivers surrounded
by grassland (tussock, pasture). Consistent with our hypothesis, food webs
of rivers flowing through grassland, with little detritus input, and thus
with substantial environmental pressure,
are more hierarchical than food webs of rivers flowing through forests,
which provide greater detritus input.

\begin{figure}[htbp]
    \begin{center}
        \includegraphics[scale = 0.8]{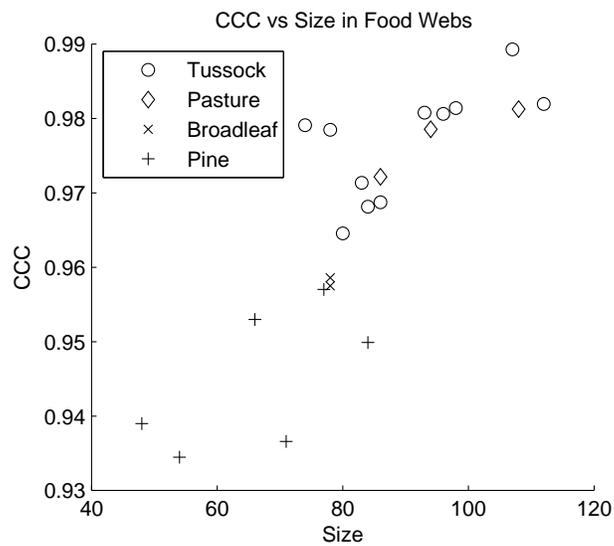}
    \end{center}
    \caption{The studied food webs show a trend between network size and CCC. Data from \cite{Thompson2003, Thompson2005}.}
    \label{fig:cccVsSize}
\end{figure}

\begin{figure}[htbp]
    \begin{minipage}{0.46\textwidth}
        \includegraphics[width=\textwidth]{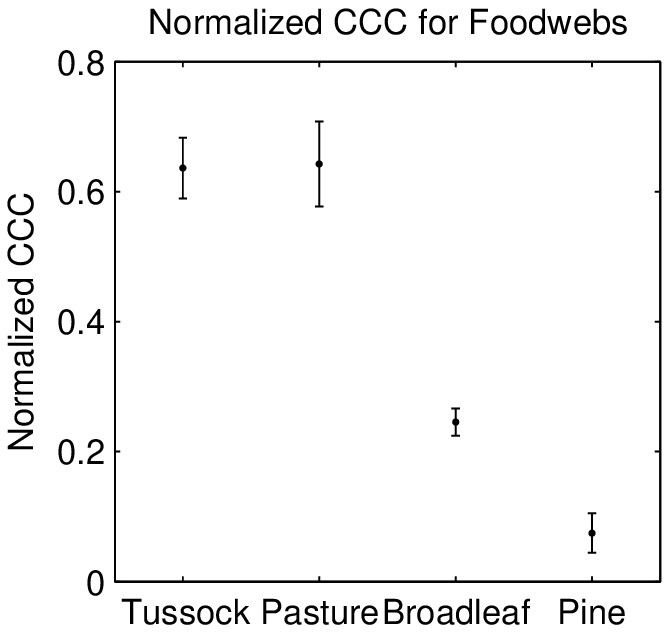}
    \end{minipage}
    \hfill
    \begin{minipage}{0.46\textwidth}
        \includegraphics[width=\textwidth]{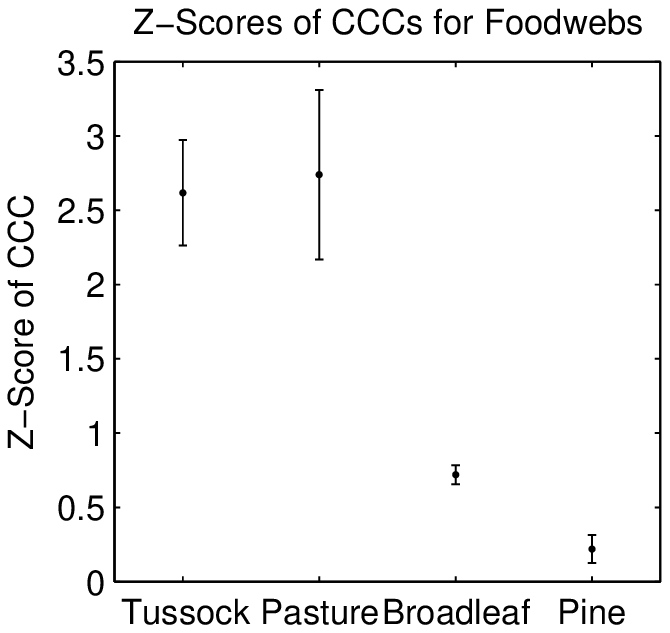}
    \end{minipage}
    \caption{Normalized CCC values and standard scores (z-scores) of river food webs flowing through different habitats relative to the distribution of CCC values obtained from random networks of equal size and total number of edges. The graphs show the same trend as the unnormalized CCC values, demonstrating that the trend is not an artifact caused by variations in network size.
The error bars are one standard error.
 Data from \cite{Thompson2003, Thompson2005}.}
    \label{fig:cccNorm}
\end{figure}

In the particular data set under study, we observed a trend between CCC values and network size that is shown in figure \ref{fig:cccVsSize}. To determine whether this trend may bias our results we compared the CCC values of each food web to the distribution of CCC values of the largest connected components of 100 random networks of the same size and total number of edges as the food web. This allowed us, in analogy to the normalized modularity given in \cite{Parter2007}, to define a normalized CCC as
\begin{align}
    \operatorname{CCC}_{\mathrm{norm}} = \frac{\operatorname{CCC} - \operatorname{CCC}_{\mathrm{rand}}}{1 - \operatorname{CCC}_{\mathrm{rand}}},
\end{align}
where $\operatorname{CCC}$ is the CCC value of the food web, and $\operatorname{CCC}_{\mathrm{rand}}$ is the average CCC value of the random networks. To consider not only the mean of the CCC values of the random networks, but also their distribution, we also computed the standard score (z-score) of each food web CCC relative to the distribution of CCC values of the random networks of the same size and sparsity:
\begin{align}
    Z_{\mathrm{CCC}} = \frac{\operatorname{CCC} - \operatorname{CCC}_{\mathrm{rand}}}{\sigma},
\end{align}
where $\sigma$ is the standard deviation of the CCC values of the random networks. As shown in figure \ref{fig:cccNorm}, the normalized CCC values show the same trend as the unnormalized values, indicating that network size is not the cause of the observed trend. 

\begin{figure}[htbp]
    \begin{minipage}{0.46\textwidth}
        \includegraphics[width=\textwidth]{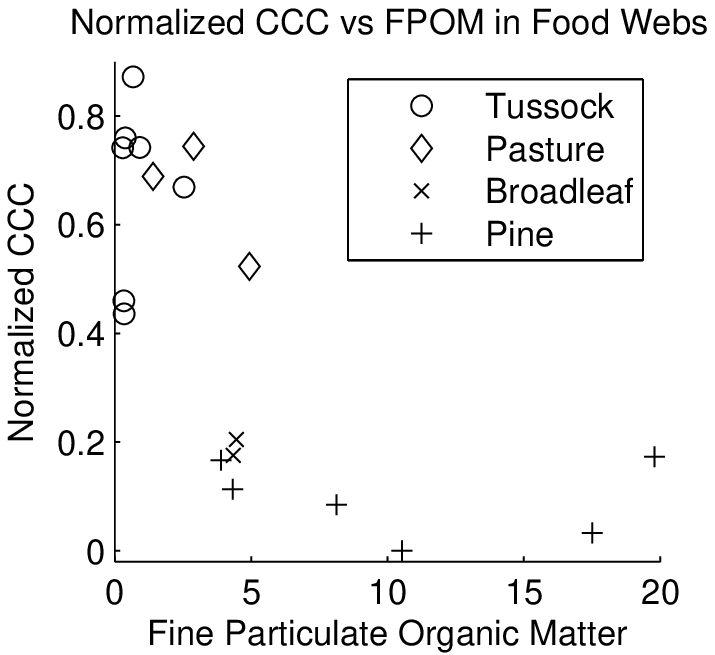}
    \end{minipage}
    \hfill
    \begin{minipage}{0.46\textwidth}
        \includegraphics[width=\textwidth]{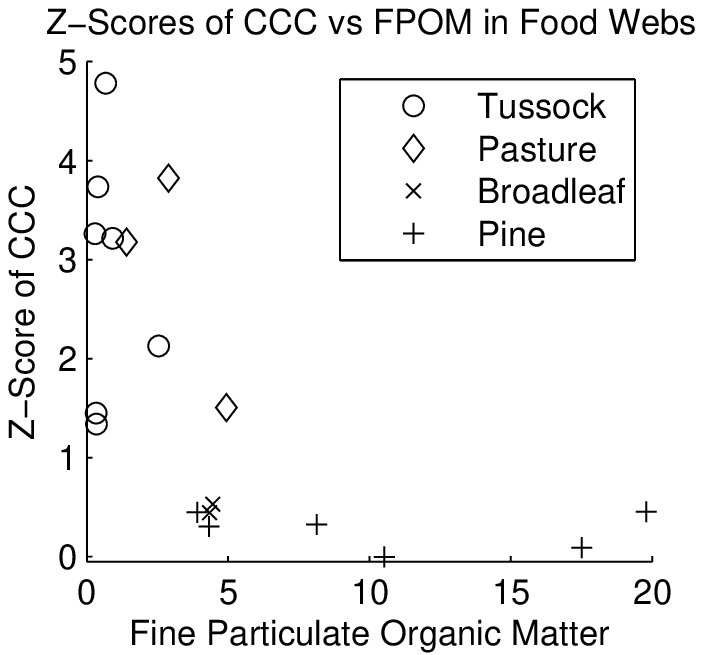}
    \end{minipage}
    \caption{Normalized CCC values and standard scores (z-scores) of river food webs as a function of fine particulate organic matter. Both normalized CCC measures decrease with decreasing particulate organic matter. Data from \cite{Thompson2005}.}
    \label{fig:cccVsFPOM}
\end{figure}

For most of the food webs that we analyzed, we were able to find explicit numbers for the amount of fine particulate organic matter present in the river from \cite{Thompson2005}. This allowed us to investigate how CCC varies explicitly with detritus input rather than with the surrounding habitat. The results in figure \ref{fig:cccVsFPOM} show a clear trend of decreasing CCC with decreasing detritus availability. Not all fine particulate organic matter in a river is a consequence of detritus input but significant detritus input will lead to more particulate organic matter. 

\begin{figure}[htbp]
    \begin{center}
        \includegraphics[scale = 0.8]{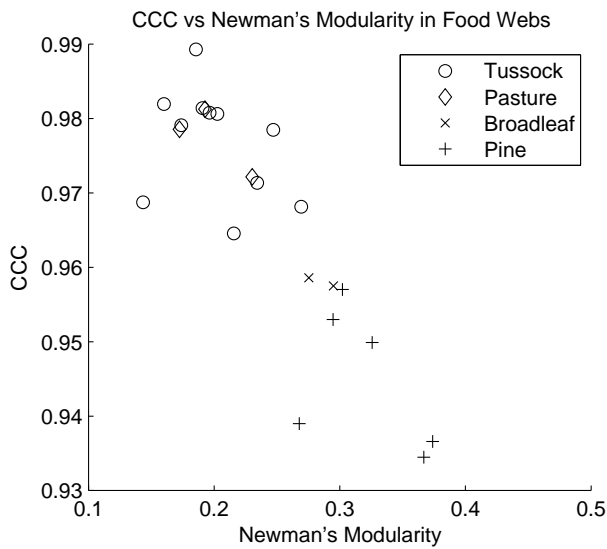}
    \end{center}
    \caption{Cophenetic correlation coefficient versus Newman's modularity for the largest connected component of river food webs. There is a negative correlation between CCC and Newman's modularity. However, the value for Newman's modularity is quite low for most of the food webs. Data from \cite{Thompson2003, Thompson2005}.}
    \label{fig:cccVsNewmanMod}
\end{figure}

The analysis of food webs also revealed that modularity and hierarchy, despite being closely related concepts, do not have to be positively correlated. Indeed, we observed the opposite trend for this food web data set. A comparison of the CCC values and Newman's modularity, maximized using the spectral algorithm described in \cite{Newman2006}, is shown in figure \ref{fig:cccVsNewmanMod}. The data show a negative correlation between these measures of hierarchy and modularity. Note also, however, that the values of Newman's modularity are quite low for most of the food webs and may not indicate significant modularity.
While we often speak of modularity as a short-hand, the fundamental 
NP $\to$ P transition is induced by multiple levels of modularity,
\emph{i.e.}\ hierarchy.  Therefore, hierarchy is the more fundamental
order parameter to consider.
Our analysis of New Zealand river food webs shows an increase in hierarchy with increasing environmental pressure.

\subsection{Modularity in Development}
It has long been recognized that the body plans of higher organisms are modular: they consist of easily identifiable parts which serve a well defined function and are structurally separated from other parts \cite{Klingenberg2008}. It has also been observed that among metazoans, modularity increases with complexity and that modules may be hierarchically structured \cite{Wagner1996}. This modularity in phenotype is an immediate consequence of a developmental modularity which can be observed over many levels from parts of genes to the scale of organisms \cite{Raff2000a}.

Modularity in development has been associated with evolvability. Raff and Sly pointed out that modularity enables the evolution of ontogeny because it makes three processes possible: the dissociation of developmental processes (\emph{e.g.}\ heterochrony), the duplication and subsequent divergence of developmental modules, and the co-option of features into new functions  \cite{Raff2000}. 
Thus, developmental modularity improves evolvability and allows for the emergence of complex anatomies from genomes which need not be as complex.
A study by Yang \cite{Yang2001} provides empirical evidence for the hypothesis that modularity confers evolvability. He compared the taxonomic diversity of insect lineages with different degrees of modularity and found that lineages with greater life-stage modularity have greater rates of diversification \cite{Yang2001}. Litvin \emph{et al}.\ showed that environmental change and intrinsic genetic variation can alter the connectivity of the modules in gene regulatory networks \cite{Litvin2009} lending additional support to the idea that modularity confers evolvability by permitting a dynamic rewiring of network components in response to environmental perturbation. That is, modularity increases evolvability.

Meir \emph{et al}.\ studied a computer model of the neurogenic network of \emph{Drosophila melanogaster} and found it to be very robust to a change in parameters or initial conditions \cite{Meir2002}. They also showed that, within their model, this robustness provides both functional and evolutionary flexibility. It allows a network performing one function to evolve the ability to achieve additional functions. In this case, robustness confers evolvability.

In 2006, Davidson and Erwin proposed that the hierarchical modular structure of gene regulatory networks leads to different rates of evolution between major aspects of body plan morphology and terminal properties of body plans \cite{Davidson2006}. They found a hierarchy with four types of modules in gene regulatory networks, ``kernels,'' ``plug-ins,'' ``switches,'' and ``batteries,'' each of which has a different function during development. Kernels shape the phylum- and superphylum- level characteristics, plug-ins and switches are associated with class, order, and family characteristics, and batteries are involved in speciation. The idea that the genetic framework upon which selection acts is not unstructured, as assumed in classic evolutionary theory, sparked a controversy \cite{Coyne2006, Erwin2006}. But a recent publication provides further evidence that the hierarchical structure of developmental regulatory networks provides an organizing structure for the evolution of the body plan
\cite{He2010}.
An explicit calculation  of the rate of evolution of genes in different types of modules in gene regulatory networks demonstrated the influence
of hierarchical structure on evolvability.

\subsection{Modularity in Physiology}
Traditionally it has been believed that the healthy physiologic state is characterized by homeostasis and that maintenance of all physiologic variables in narrow ranges around optimal values is the key to good health. Pathology on the other hand was thought to result from a deviation of one or more physiologic variables from their healthy values. As a consequence, much of Western medicine is focused on restoring physiologic variables to their normal values \cite{Buchman2002a}.

More recently this view has been questioned, and it has been suggested that variation in physiologic variables may not be a detriment to, but rather a necessary component of, health \cite{West2006}. The fluctuations observed in physiologic systems around mean values are just as important as the mean values themselves and aging and disease are characterized by a loss of variability \cite{Buchman2002a}. The hypothesis that variability is associated with health is supported experimentally by Boker \emph{et al}.\ who found that introducing noise into mechanical ventilators leads to an improvement in lung function \cite{Boker2004}.

However, not all noise is good noise. For example, atrial fibrillation, which leads to very irregular heartbeat intervals, is certainly not associated with health. Thus, the idea arose that ``complex'' physiologic time signals are indicative of healthy systems. Different measures of complexity have been suggested, most of which are based on entropy (\emph{e.g.}\ \cite{Costa2007}). Here we propose that complexity in physiology can be understood as modularity, and that modularity deteriorates in aging and disease. Healthy physiology is characterized by a modular partitioning phase space that facilitates transitions between states. This modularity promotes adaptability to external stimulus. A healthy state is described neither by a complete disconnect between modules nor by very strongly connected modules. Rather, there is an optimal amount of connectedness between modules.

\subsubsection{Heart Rate}
An early study showed that the decoupling of physiologic systems is associated with a decrease in variability and with disease
 \cite{Goldstein1998}. 
Goldstein \emph{et al}.\ studied the consequences of acute brain injury on heart rate variability. They found that neurological injury leads to a decoupling of the autonomic and cardiovascular systems and that this decoupling leads to a decrease in heart rate and blood pressure variability.
It was also observed that a recoupling of cardiovascular signals is necessary for recovery.

\begin{figure}[htbp]
    \begin{center}
        \includegraphics[scale = 0.8]{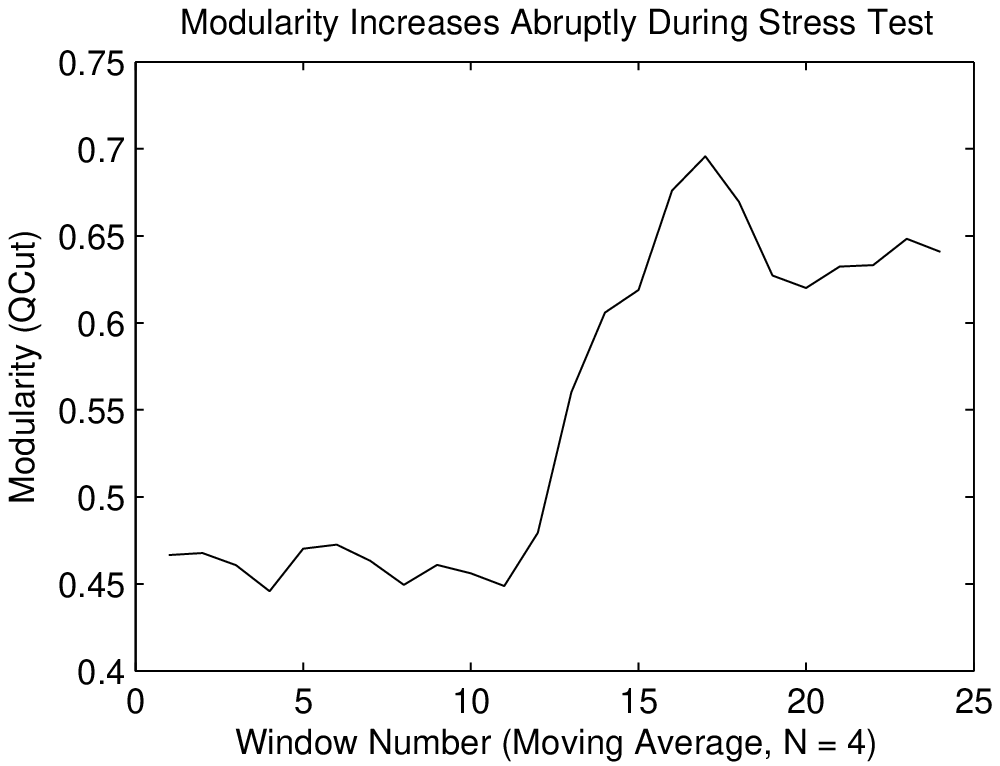}  \\
        \includegraphics[scale = 0.8]{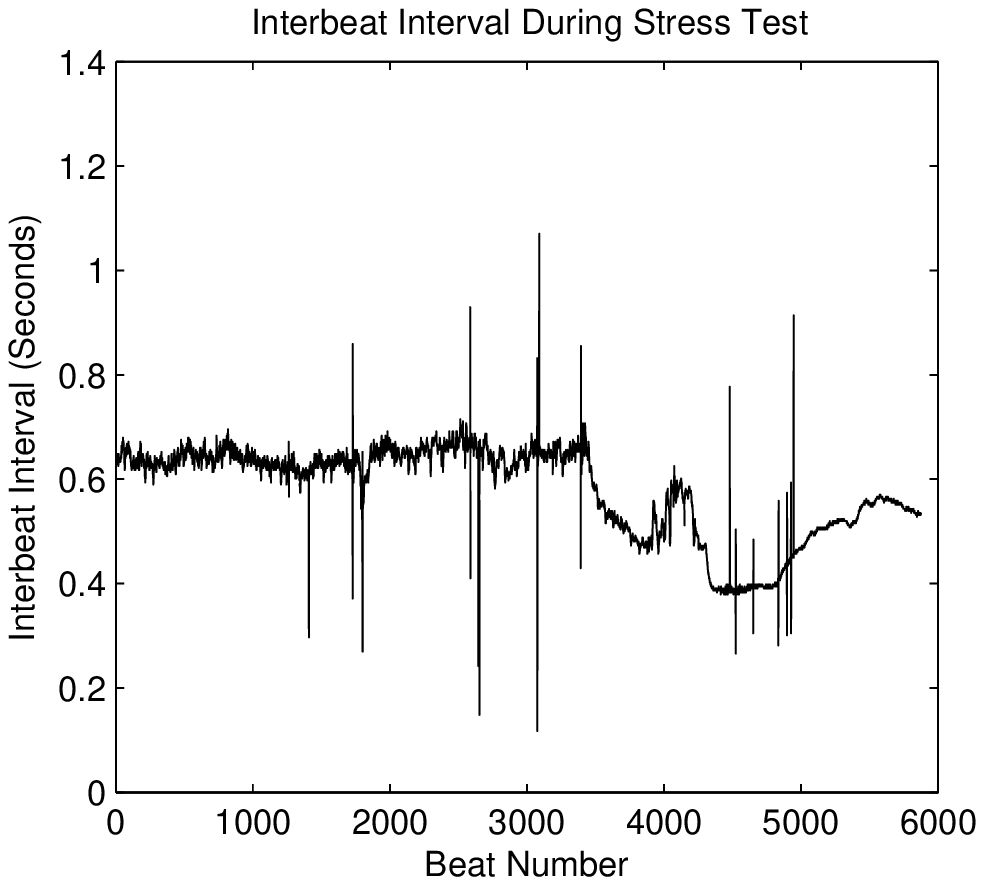}
    \end{center}
    \caption{From \cite{Buchman2010}. Upper graph: the modularity of the interbeat covariance matrix increases abruptly during a stress test on a treadmill. Lower graph: unprocessed record of interbeat intervals during the same test. The  absciss{\ae} of the two graphs correspond to the same time interval. Note that the increase in modularity precedes the rise in heart rate by 400 heart beats.}
    \label{fig:buchmanModStress}
\end{figure}
A recent study provides empirical evidence that environmental stress can increase the modularity of a physiologic system, which in turn improves the system's ability to respond to the external stress. Anton Burykin and Timothy Buchman investigated the response of the human heart beat to exercise stress tests \cite{Buchman2010}. During such a test, subjects are exposed to increasing levels of speed and inclination on a treadmill. After some time for acclimatization in each level, the exercise load is increased. When the subject reaches maximal exertion, exercise load is decreased. Burykin and Buchman analyzed the obtained heart beat time series data by constructing an interbeat covariance matrix and computing the modularity of this matrix. They observed, as shown in figure \ref{fig:buchmanModStress}, that the modularity increases abruptly in response to the external stress. It is interesting to note that the modularity already begins to increase 400 heart beats before a change in the heart beat frequency is observed \cite{Buchman2010}.

\begin{figure}[htbp]
    \begin{center}
        \begin{minipage}{0.3\textwidth}
            \begin{center}
                \includegraphics[width = \textwidth]{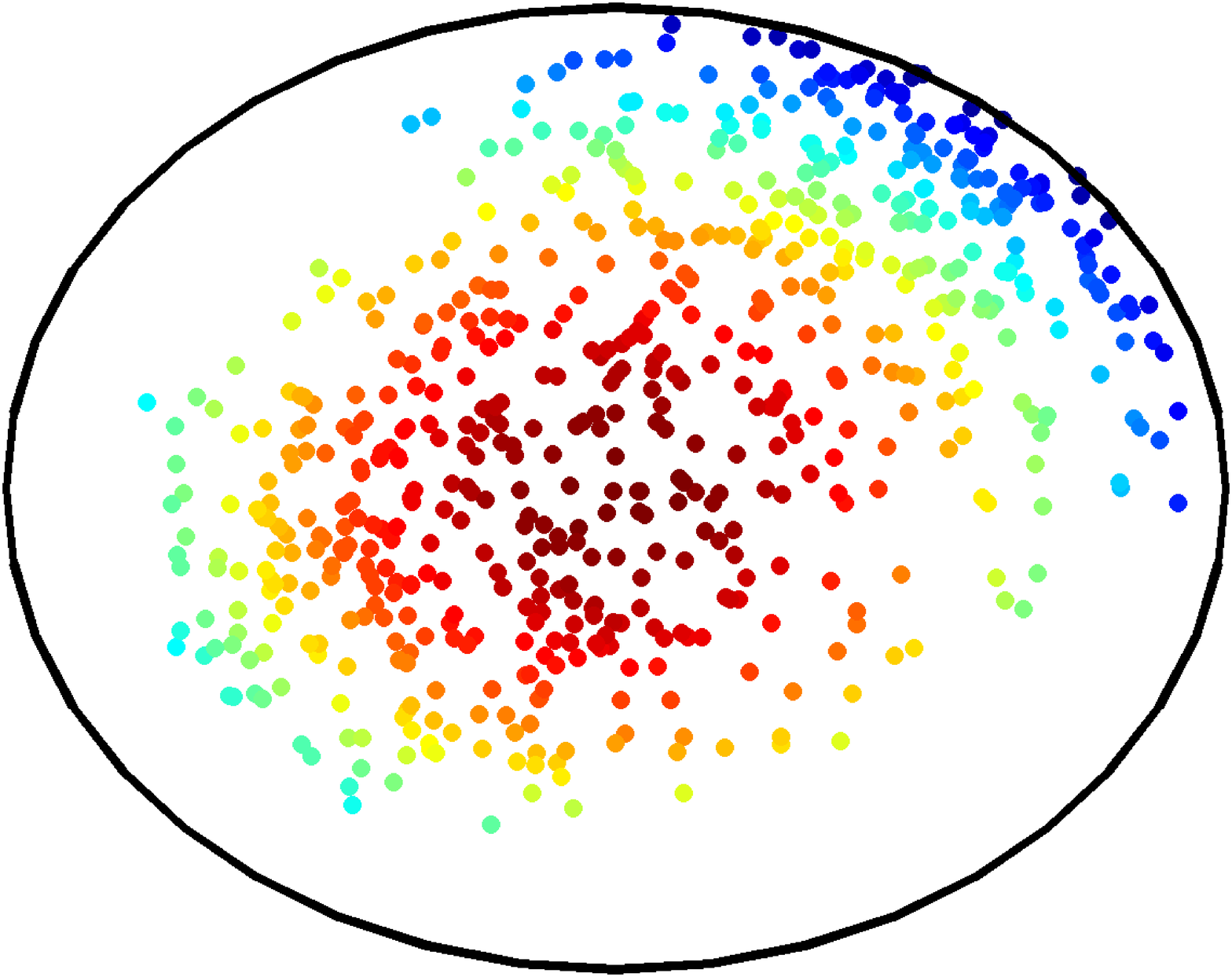} \\ (a) AF N=6 \\
                \includegraphics[width = \textwidth]{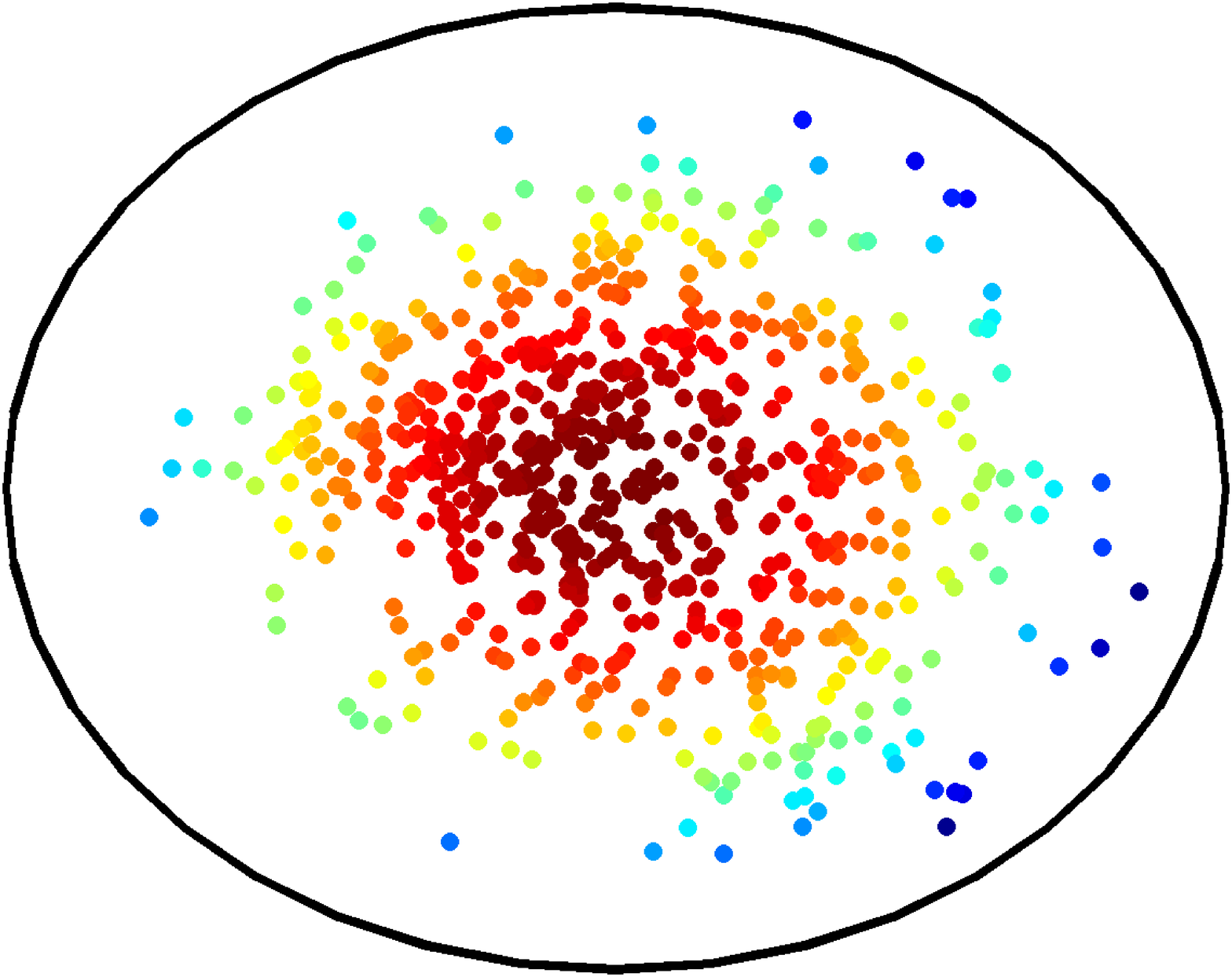} \\ (d) AF N=11 \\
                \includegraphics[width = \textwidth]{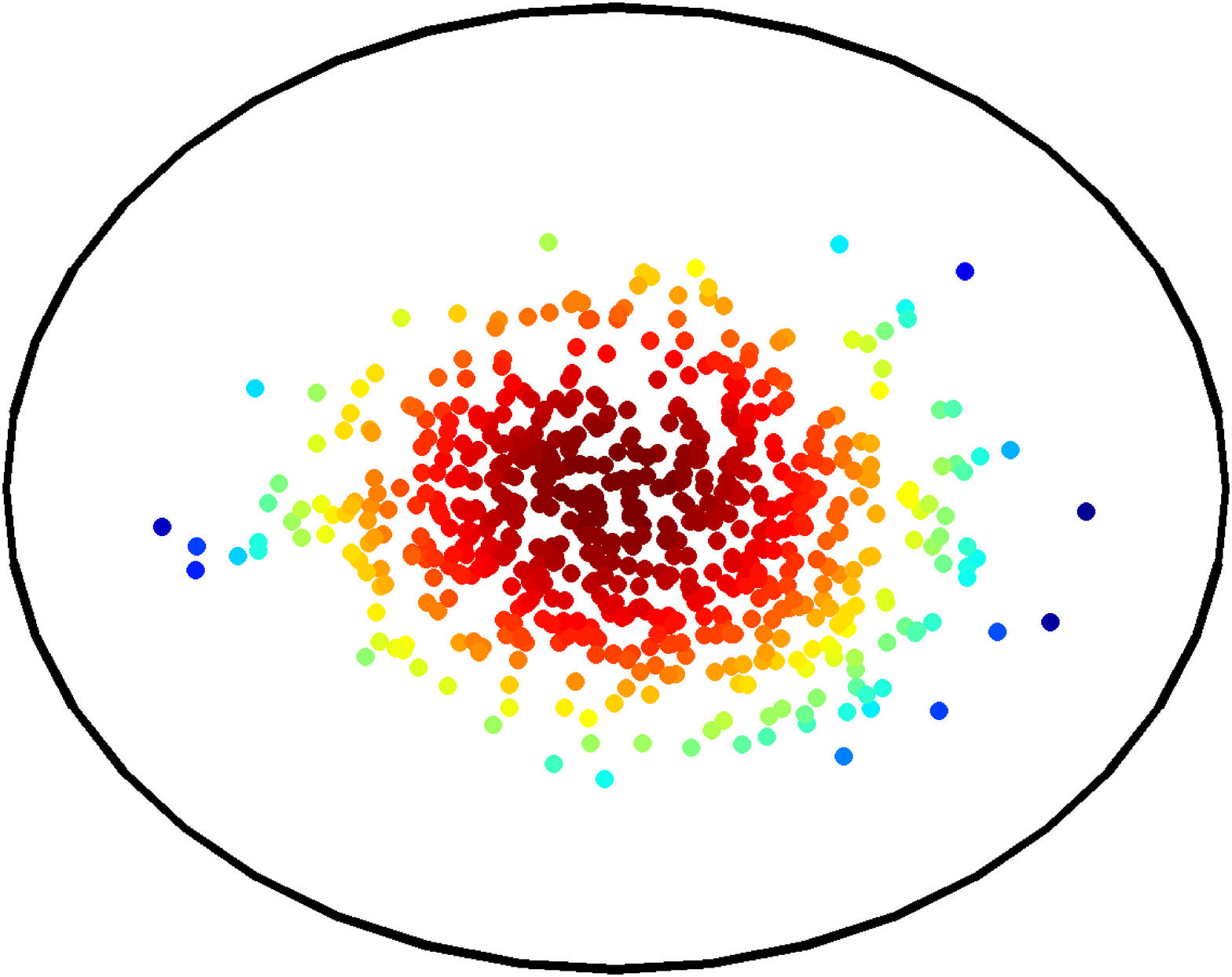} \\ (g) AF N=16
            \end{center}
        \end{minipage}
        \begin{minipage}{0.3\textwidth}
            \begin{center}
                \includegraphics[width = \textwidth]{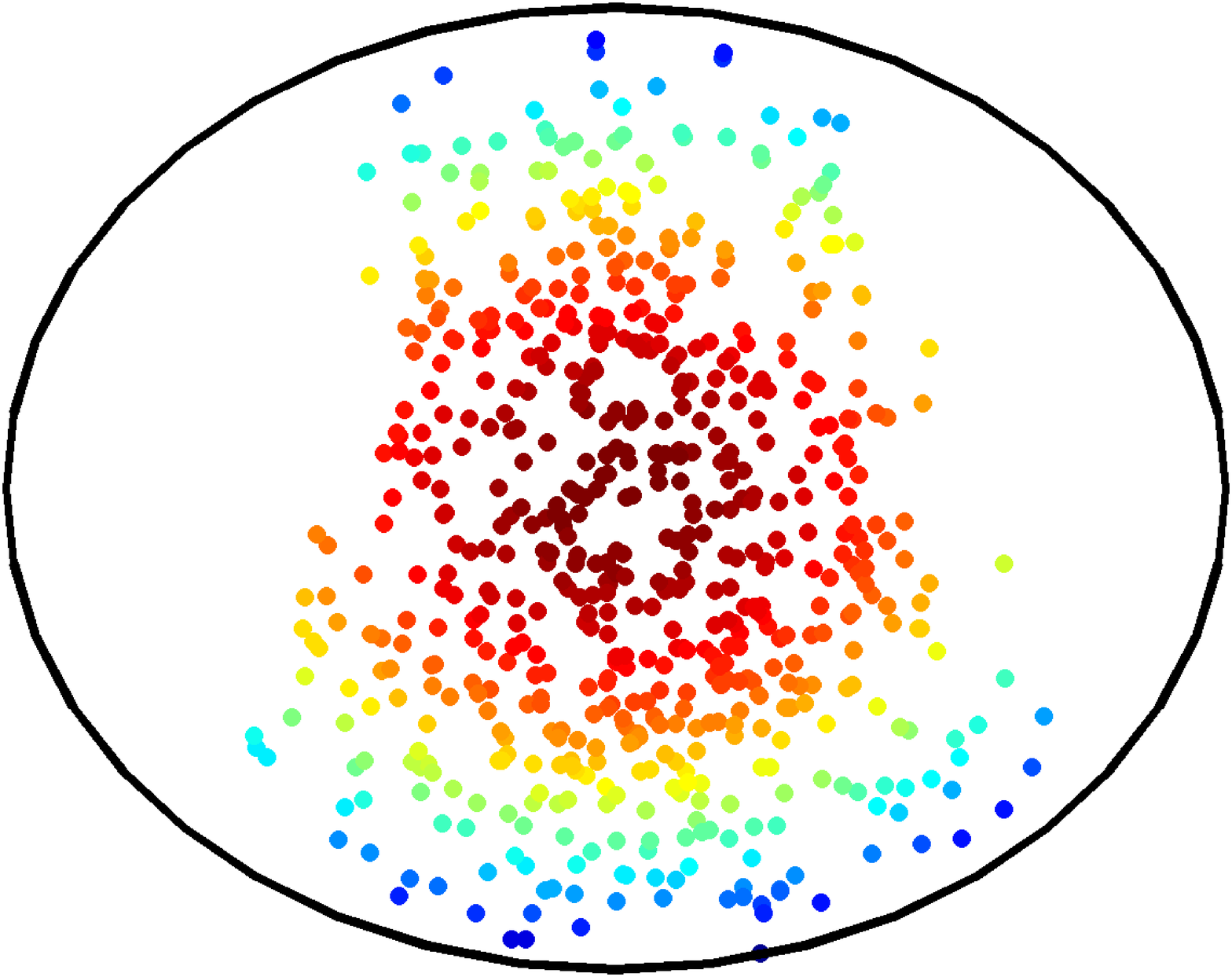} \\ (b) CHF N=6 \\
                \includegraphics[width = \textwidth]{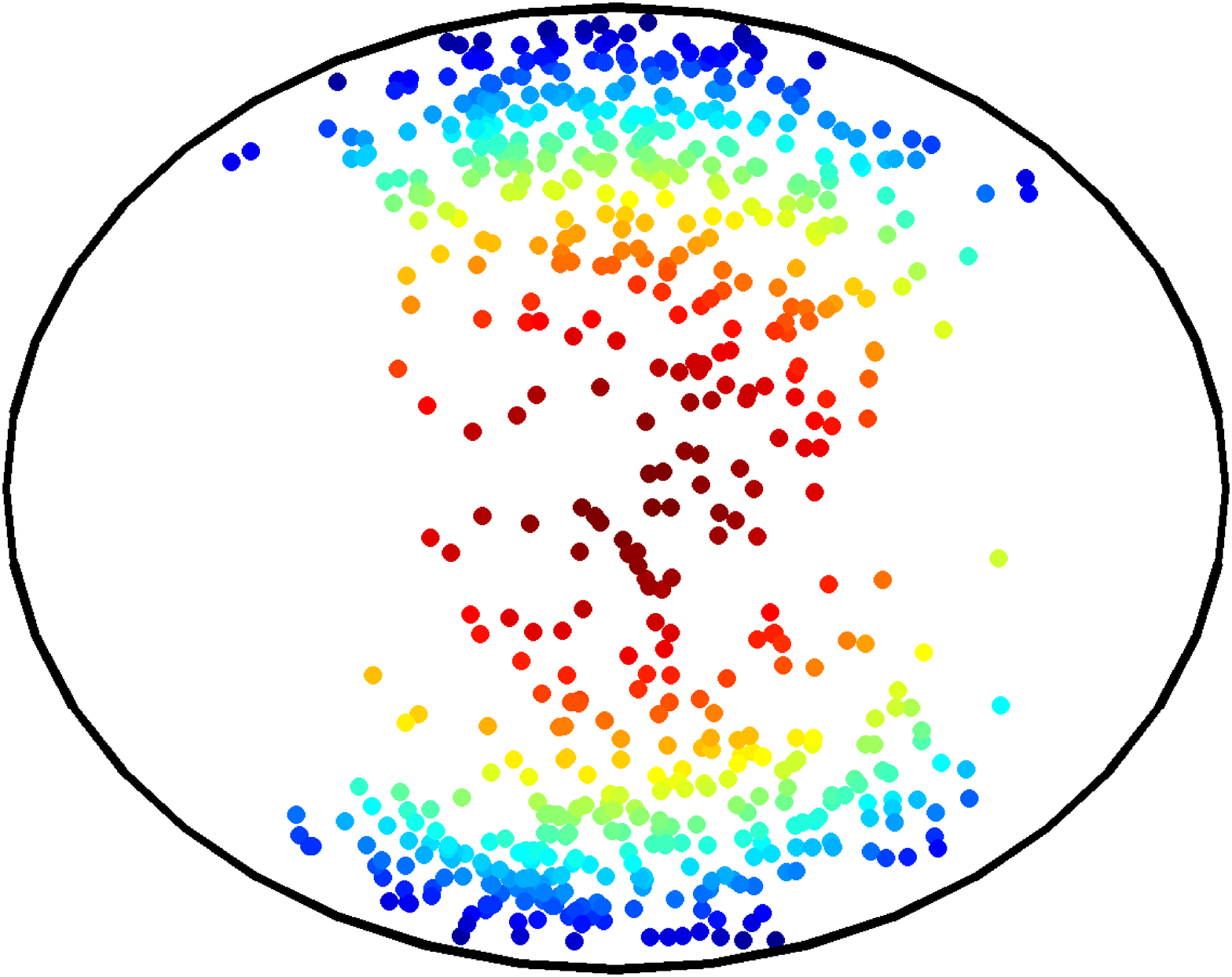} \\ (e) CHF N=11 \\
                \includegraphics[width = \textwidth]{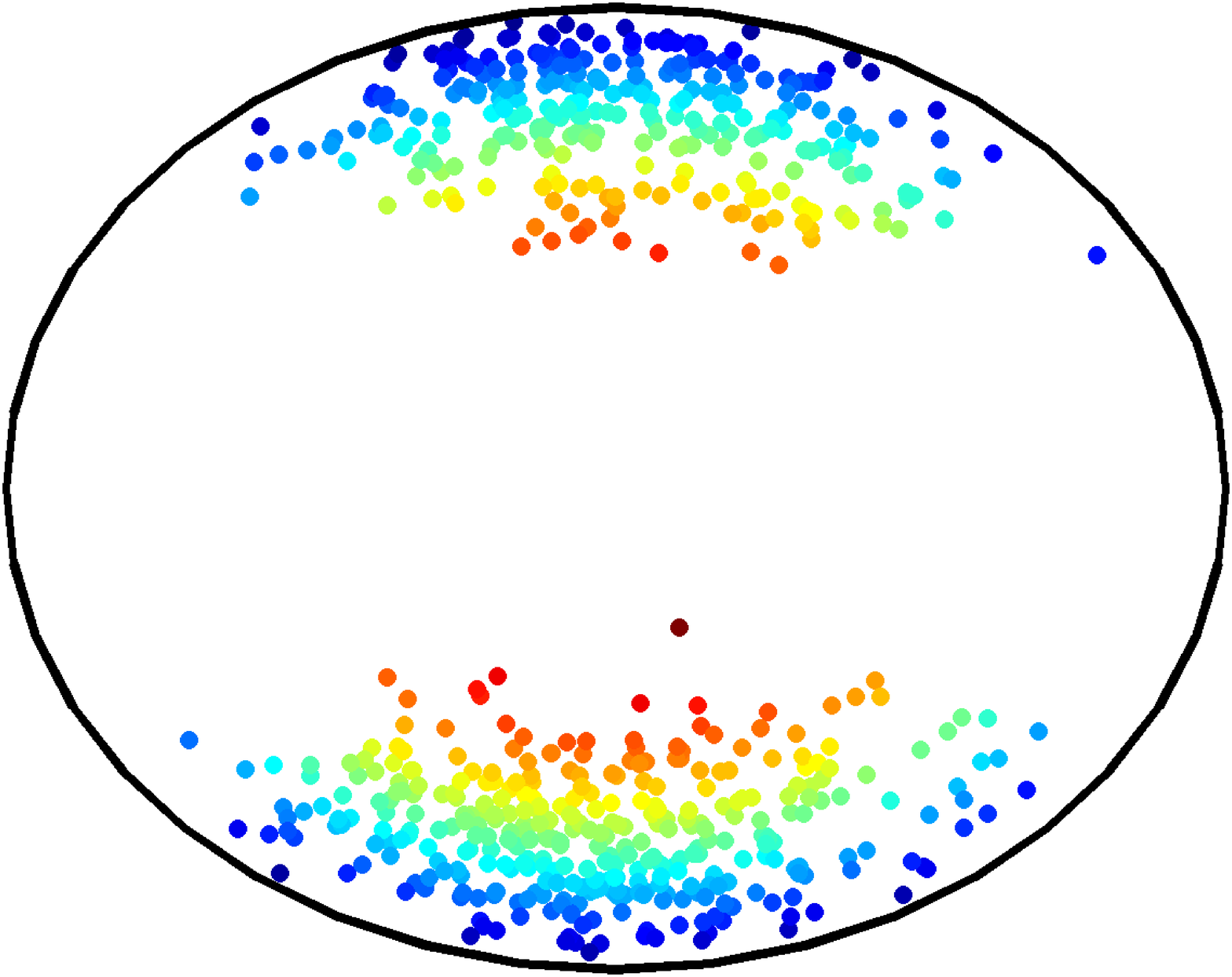} \\ (h) CHF N=16
            \end{center}
        \end{minipage}
        \begin{minipage}{0.3\textwidth}
            \begin{center}
                \includegraphics[width = \textwidth]{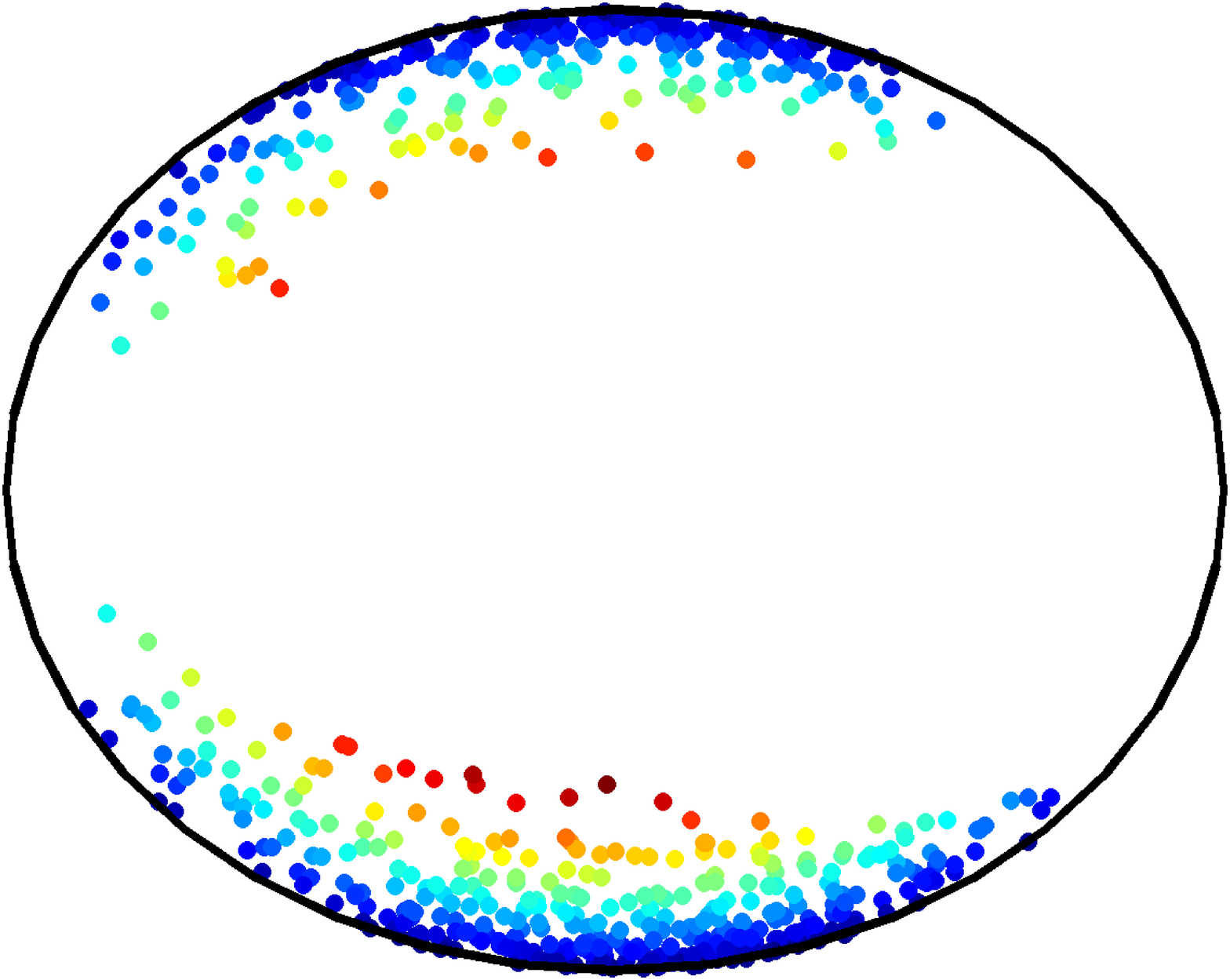} \\ (c) HL N=6 \\
                \includegraphics[width = \textwidth]{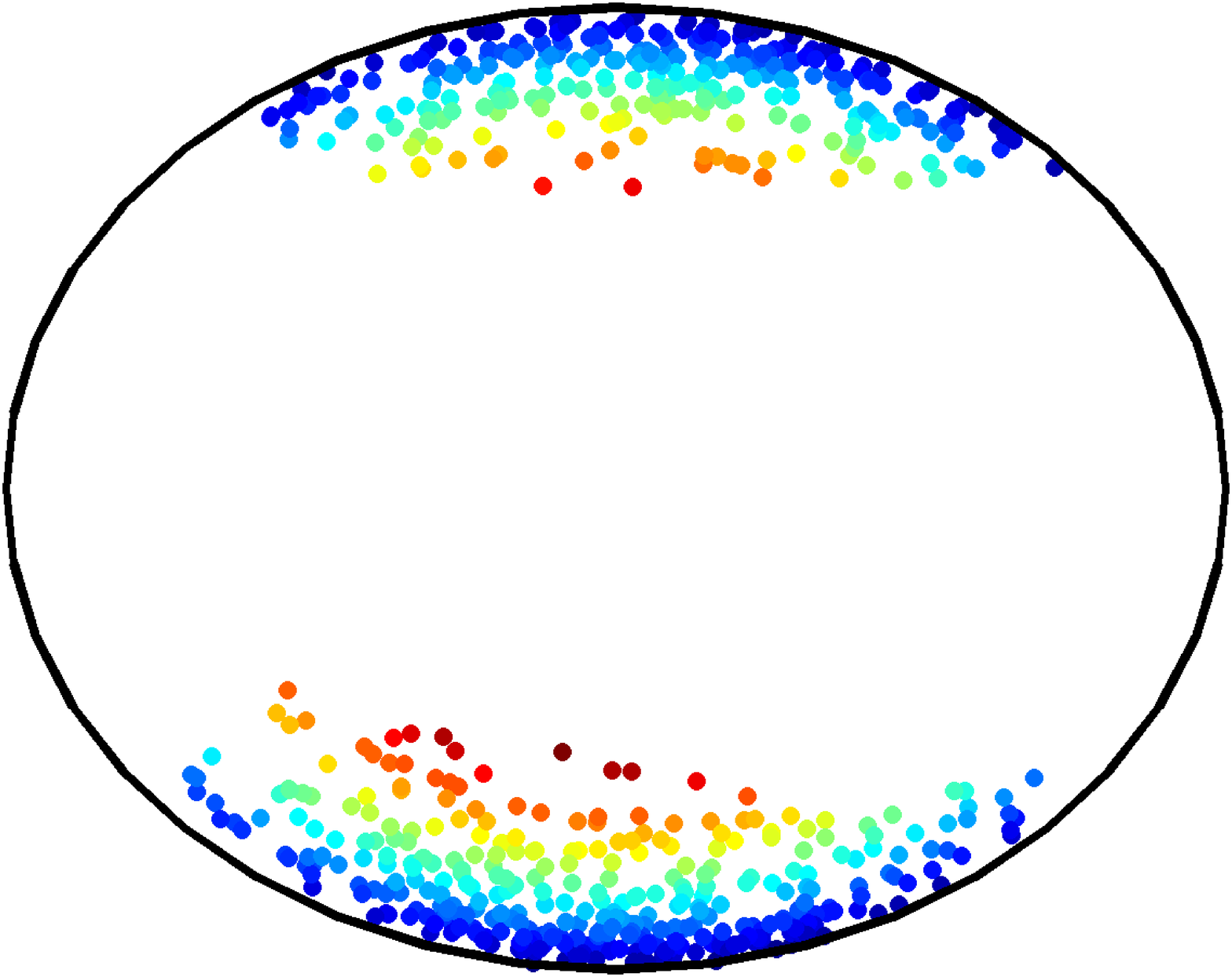} \\ (f) HL N=11 \\
                \includegraphics[width = \textwidth]{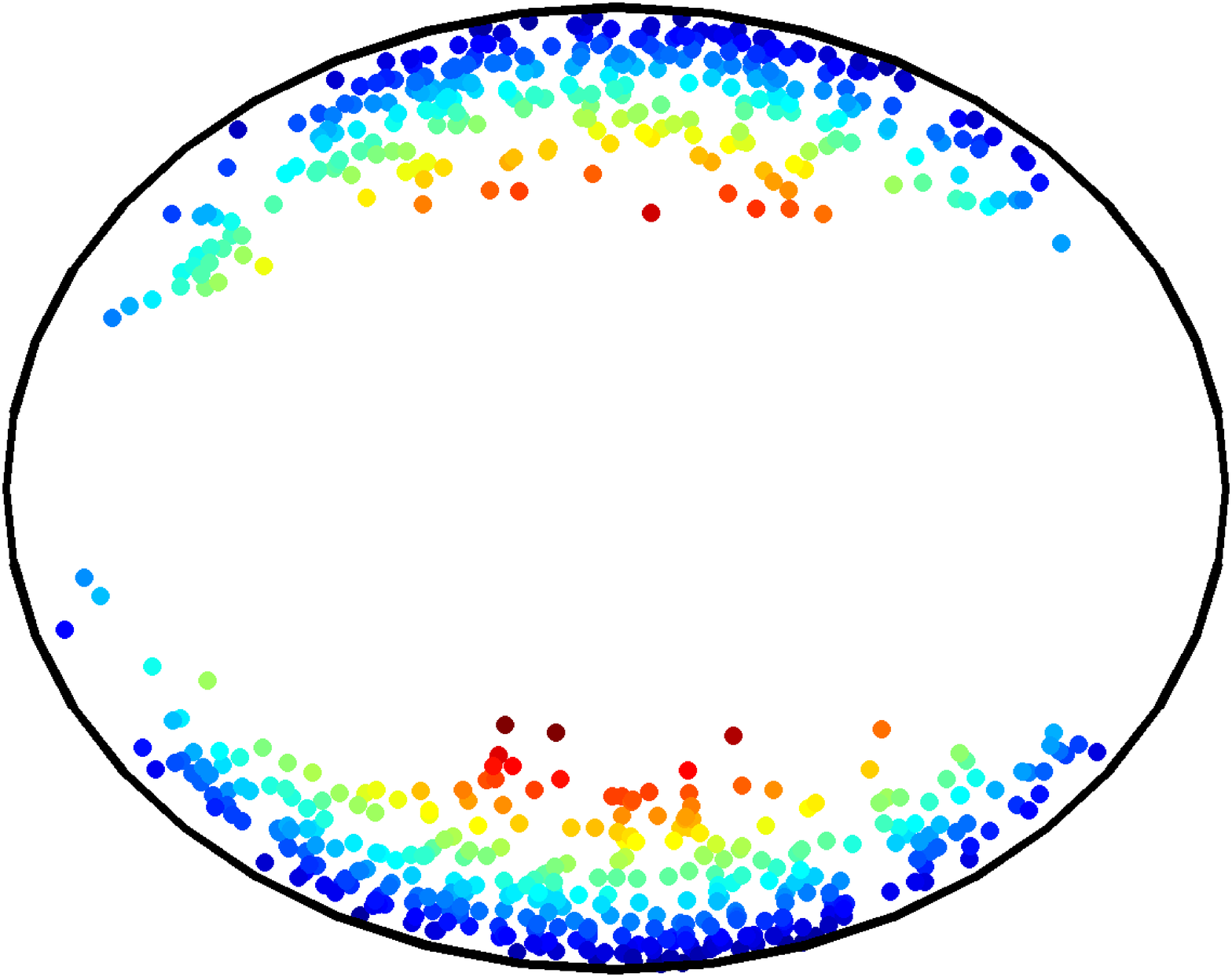} \\ (i) HL N=16
            \end{center}
        \end{minipage}
    \end{center}
    \caption{Results of a topological analysis of frequent-motif spaces performed by Carlsson \emph{et al}.\ \cite{Carlsson2010} for different time lengths, $N$. The motifs of healthy patients (HL) always separate into two components while the motifs of patients suffering from atrial fibrillation (AF) never show a separation. For congestive heart failure (CHF) patients, the motif space shows no separation for small $N$, but
a separation develops
 with increasing $N$.
}
    \label{fig:carlsson}
\end{figure}

Similarly, an experiment conducted by Carlsson \emph{et al}.\ supports the hypothesis that disease is associated with a decrease in modularity
 \cite{Carlsson2010}. 
 The researchers compared heart interbeat interval time series data from three groups: healthy subjects, patients suffering from atrial fibrillation (AF), and patients suffering from congestive heart failure (CHF).
They extracted motifs that occurred frequently in each patient's data set and performed a topological analysis of the space of these motifs. Their results, shown in figure \ref{fig:carlsson}, show the presence of two clear modules in the space of frequent motifs of healthy subjects that can be observed for all time scales. The motif space of AF patients does not separate into modules, while the motif space of CHF patients develops a separation only for longer time scales \cite{Carlsson2010}.

\subsubsection{Postural Control}
The benefits of explicitly introducing variability to restore the functionality of a physiologic system has been observed. 
In the context of postural control, Collins \emph{et al}.\ showed that noise can enhance the detection of a subthreshold tactile stimulus \cite{Collins1996}, a phenomenon referred to as stochastic resonance. To test the concept of stochastic resonance in postural control, Priplata and coworkers applied subsensory noise to the feet of young and elderly subjects during quiet standing
 \cite{Priplata2002, Priplata2003}.
This noise resulted in a reduction of postural sway in both groups, with a larger improvement in the elderly.
 In a follow-up study, they extended the group of subjects to include patients with diabetes and patients who had had a stroke. Subsensory noise led to improved balance in all groups; the improvement was greater for subjects with greater baseline sway (worse balance) \cite{Priplata2006}. Costa \emph{et al}.\ quantified complexity by a measure called multiscale entropy and observed that the complexity of postural sway dynamics in elderly subjects with a history of falls is lower compared to that of both young subjects as well as elderly subjects without a history of falls
 \cite{Costa2007}. 
 Applying subsensory noise to the feet increased the complexity of sway fluctuations in the elderly.
These results show that variability is an essential part of healthy physiology.

\subsubsection{Brain Networks}
\begin{figure}[htbp]
    \begin{minipage}{0.46\textwidth}
        \includegraphics[width=\textwidth]{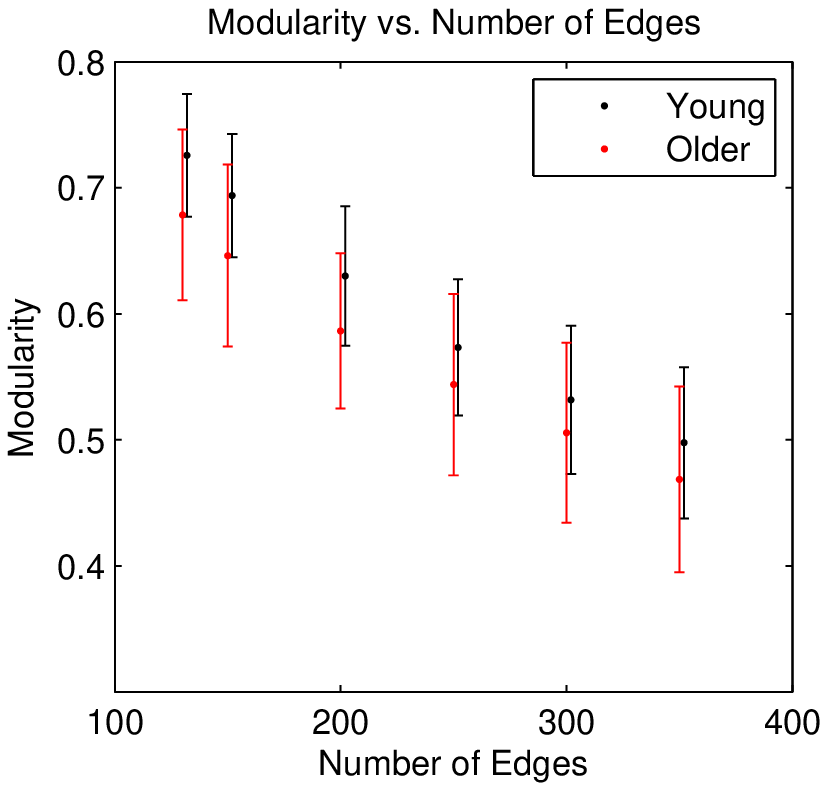}
        \caption{Newman's modularity for different threshold values, following 
\cite{Meunier2009}. 
The modularity in the younger group is consistently larger than in the older group. The difference becomes more significant for larger threshold values, \emph{i.e.}\ sparser networks. The modularities shown here are slightly larger that in
\cite{Meunier2009}
because we used the spectral algorithm described in \cite{Newman2006} to maximize modularity rather than a greedy algorithm.}
        \label{fig:modYoungOlderVsEdges}
    \end{minipage}
    \hfill
    \begin{minipage}{0.46\textwidth}
        \includegraphics[width=\textwidth]{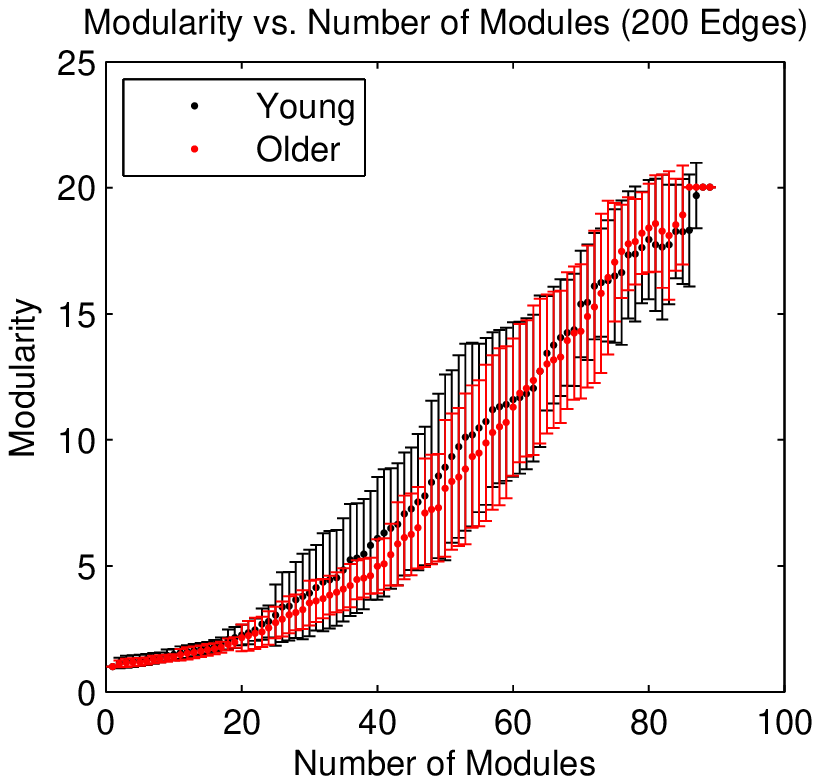}
        \caption{The modularity as defined in equation \eqref{eq:modularity} across the dendrogram obtained from average linkage hierarchical clustering on commute distance matrices of network divisions for 200 edges. The modularity of brain networks from young adults is higher than that of networks from older adults for all relevant parts of the dendrogram. Data from \cite{Meunier2009}.}
        \label{fig:modYoungOlder200}
    \end{minipage}
\end{figure}

A relation between modularity on the one hand and aging and pathology on the other, which is similar to the one in cardiovascular signals,
 has also been observed in brain networks. Meunier \emph{et al}.\ studied the modular structure of human brain functional networks in young and older adults and found that both showed significant modularity and that the network structure of the human brain changes with age \cite{Meunier2009}. They also displayed the modularity of the two groups as a function of the number of edges,
\emph{i.e.}\ applied different threshold values, and observed that the modularity of the young group is consistently higher than the modularity of the older group. This difference was not statistically
significant for any of the cutoff values used, but it is apparent in the graph that the difference becomes more significant as the number of edges decreases. We extended the analysis to networks with 130 edges, slightly below the lowest value of 150 used by Meunier \emph{et al}., and observed a 
statistically significantly higher modularity in the younger group than in the older group. Our results are shown in figure \ref{fig:modYoungOlderVsEdges}.

To further support the hypothesis that brain networks of young adults are more modular than those of older adults, we used a completely different method to quantify modularity on the same networks. We computed a matrix of Euclidean commute time distances, as described in section \ref{sec:modularityInEcology} for each brain network and used average linkage hierarchical clustering to create a dendrogram of possible partitions for each network. Then we quantified modularity as the ratio of intramodule weight over intramodule area in the adjacency matrix of the network \cite{He2009}:
\begin{align}
    M = \frac{\displaystyle\sum_{j,k \ne j} A_{jk} \delta(c_j, c_k)}{\displaystyle\sum_{j,k \ne j} A_{jk}}
            \times \left(\frac{\displaystyle\sum_{j,k \ne j} \delta(c_j, c_k)}{n(n - 1)}\right)^{-1},
    \label{eq:modularity}
\end{align}
where $A_{jk}$ is the adjacency matrix of the network and $\delta(c_j, c_k) = 1$ if nodes $j$ and $k$ are in the same module and $\delta(c_j, c_k) = 0$ otherwise. The results, shown in figure \ref{fig:modYoungOlder200}, show that the modularity of brain networks from young adults is consistently higher than that of brain networks from older adults across the relevant part of the dendrogram of network divisions.

Empirical evidence also shows that the brain networks of diseased patients are less modular than those of healthy subjects. Chavez \emph{et al}.\ analyzed the structure of brain networks from magnetoencephalographic signals in epileptic patients and compared it to that of healthy controls
 \cite{Chavez2010}. 
They observed that the patients suffering from epilepsy had brain networks with greater connectivity and lower modularity than the brain networks of the controls.
They also found that in epilepsy patients, nodes have more connections to nodes in different functional modules.

Studies on very different physiological systems reveal a common trend: aging and disease lead to a decrease in the modularity of physiological systems, which reduces the ability of these systems to respond to external stress.

\subsection{Social Networks}
The emergence of hierarchical structure in response to environmental pressure has also been observed in social networks. Social networks
 are evolving systems. Thus, the insights gained from the study of general evolving systems may be applied to understand their behavior. In a very recent publication \cite{He2010a}, the temporal evolution of hierarchical structure in the world trade network was analyzed over the last 40 years. It was shown that during recessions, which can be considered a form of environmental pressure, the world trade network tends to become more hierarchical, with a larger observed increase in hierarchy during more severe recessions. In addition it was found that globalization transforms the trade network into a less hierarchical state. This decreased hierarchy makes the trade network more sensitive to environmental shocks and leads to a slower recovery after recessions. These observations are again consistent with the theory presented in section \ref{sec:spontaneousEmergence}. Increased environmental pressure leads to hierarchical structures which, in turn, improve a system's ability to respond to environmental perturbations.

\section{Conclusion}
\begin{figure}[htbp]
    \begin{center}
        \includegraphics[width = 0.6\textwidth]{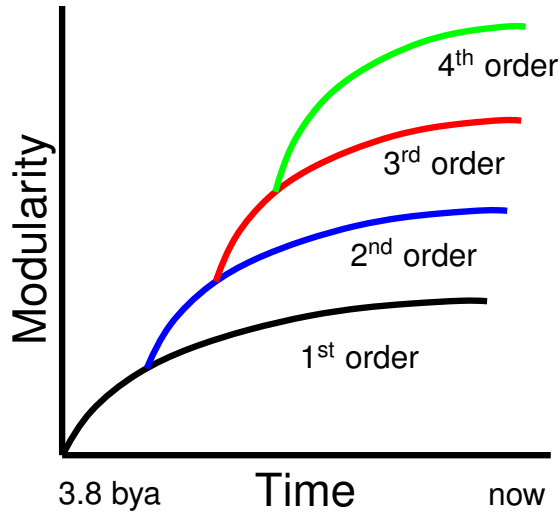}
    \end{center}
    \caption{The emergence of hierarchical modularity. Initially, modularity increases as first-order modules are formed. However, modularity is bounded,
 and so there is a time at which it saturates. We expect that before 
first-order modularity saturates,  a second-order modularity emerges by combining the first-order modules into second-order modules. This process repeats to create a hierarchy of modules.}
    \label{fig:higherOrderMod}
\end{figure}
We have presented the hypothesis
that a changing environment selects for adaptable frameworks, and
competition among different evolutionary frameworks leads to
selection of structures with the most efficient dynamics, which are
the modular ones.
From a computer science point of view, by forming a hierarchy, the NP-complete problem of searching the entire sequence space is replaced by a polynomial-time approximation.    Many low-lying states are lost, but those that remain are found more quickly.
From a physics point of view, the Hamiltonian is being made somewhat separable.  Shorter modules are exponentially more easily evolved.
Natural (and man-made) systems were shown in several
examples to employ modularity to a non-zero extent.

We have defined a module to be a component that can operate relatively independently of the rest of the system. Modularity was said to have emerged when there are more intramodule connections than intermodule connections. 
We reviewed the hypothesis that modularity arises because there is a generic requirement
for a system in a changing environment to be evolvable. 
This theory of spontaneous emergence of modularity
states that systems become modular under three conditions: 
changing environments, information exchange, and slow evolution.  
These conditions appear to be met in much of biological evolution.

Mathematically, modularity measures the compartmentalization of
biological organization.  In the form of a linear expansion,
the theory of spontaneous emergence of modularity 
would be stated as the rate of change of modularity 
is proportional to the environmental change:
\begin{equation}
p_{\rm E} - p_0 = \frac{M'}{R}
\end{equation}
where $p_{\rm E}$ is the environmental pressure, 
$R$ is the resistance to evolution or ruggedness of the 
fitness landscape, $M'$ is the rate of change of 
modularity, and $p_0$ is the initial value of environmental
pressure for which the system had been in steady state.
The theory reviewed here explains how 
environmental pressure, horizontal gene transfer rate, and ruggedness
of the fitness landscape promote the emergence of modularity.  This theory
was shown to explain features on scales ranging from proteins to physiology
to social networks. 

Additional theoretical challenges lie in explaining how multiple levels
of hierarchy form.  One idea is that as the benefit from the development
of modularity at one level saturates, an additional, higher-order level
is nucleated, figure \ref{fig:higherOrderMod}.
Numerical support for this idea was shown in figure \ref{fig:heModularity}.
Mathematical description of this hierarchical partitioning of
biological space would seem to be an interesting, open
research topic.

\section{Acknowledgments}
We would like to thank Timothy Buchman, Gunnar Carlsson, and Jiankui He for providing us with unpublished results and Ed Bullmore for sharing data with us. We would also like to thank all members of the FunBio team for stimulating discussions and two anonymous referees for their insightful comments and helpful suggestions which improved this paper.  This work was supported by DARPA
grant \#HR0011--09--1--0055.


\appendix


\section{Quantitative Definitions of Modularity and Hierarchy}
In this appendix we will summarize the quantitative definitions of modularity and hierarchy used throughout this review paper. A recent and thorough exposition of definitions of modularity and algorithms for the detection of modules can be found in \cite{Fortunato2010}.

\subsection{Newman's Modularity}
Newman's modularity is one of the most widely used quantitative measures of modularity. 
Consider an undirected graph with adjacency matrix $A_{ij}$ and a partition of this graph into clusters or modules defined by $\{c_i\}$, where $c_i$ describes which module node $i$ belongs to. Then, Newman's modularity is defined as \cite{Newman2006}
\begin{align}
    Q = \frac{1}{2m} \sum_{i,j} \left(A_{ij} - \frac{k_i k_j}{2m}\right) \delta(c_i, c_j),
\end{align}
where $k_i = \sum_j A_{ij}$ is the degree of each node, $m = \frac12 \sum_i k_i = \frac12 \sum_{ij} A_{ij}$ is the total number of edges and $\delta(c_i, c_j)$ is defined as
\begin{align}
    \delta(c_i, c_j) = \begin{cases} 1 &\quad\text{if nodes $i$ and $j$ belong to the same module} \\ 0 &\quad\text{otherwise.} \end{cases}
\end{align}

Conceptually, Newman's modularity compares the fraction of within-module edges in the graph to the expected fraction of within-module edges in a random graph with the same degree sequence as given by the configuration model. The value of $Q$ is normalized such that it always lies in the interval $(-1,1)$. The definition of $Q$ has also been adapted to bipartite networks \cite{Barber2007} and directed networks \cite{Leicht2008} by choosing a different null model.

The value of $Q$ depends not only on the graph under investigation but also on the chosen partition. Thus, it is not a property of the graph itself. However, numerous methods, usually for the purpose of community detection, have been developed to search for the partition of the graph that will maximize $Q$. This maximal $Q$ is a property of the graph and can be considered as a measure of modularity.

Although the method of maximizing $Q$ is widely used for community detection, some shortcomings have been identified \cite{Good2010, Fortunato2010}. First, it has been recognized that modularity maximization can fail to detect modules which are very small relative to the size of the network even if these modules are clearly defined. This effect is referred to as the resolution limit. Second, there are exponentially many partitions of the network, which can greatly differ from one another, with modularity scores that are similar to the maximum modularity. Good \emph{et al.} call this phenomenon the degeneracy problem. Finally, the maximum modularity of a network can have a strong dependence on the size of the network and the number of modules, which complicates the comparison of modularity values between different networks. Some of these problems are more severe in networks which are sparse or hierarchical as many biological networks are. While the first two of these three problems can make the identification of modules more difficult, they do not seriously affect efforts to quantify modularity in networks. The third of these problems, however, has to be addressed when studying the effect a variable may have on modularity in biological networks. One approach is to normalize the modularity of a network by comparing it to a distribution of random networks which share the same topological features such as network size and degree distribution, as it was done in \cite{Parter2007}:
\begin{align}
    Q_{m} = \frac{Q_\mathrm{real} -Q_{\mathrm{rand}}}{Q_\mathrm{max} - Q_{\mathrm{rand}}}.
\end{align}
Here $Q_{m}$ is the normalized modularity, $Q_\mathrm{real}$ is the raw modularity, $Q_\mathrm{rand}$ is the average modularity of the random networks, and $Q_\mathrm{max}$ is the upper bound of the modularity, which can either be estimated \cite{Parter2007} or taken to be the largest value from the distribution of random networks.

\subsection{Other Measures of Modularity}
A further quantitative measure of modularity given a network with adjacency matrix $A_{jk}$ and a partition of this network into modules is the ratio of the fraction of the weight within modules (\emph{coverage}) over the fraction of the off-diagonal area within modules \cite{He2009}:
\begin{align}
    M = \frac{\displaystyle\sum_{j,k \ne j} A_{jk} \delta(c_j, c_k)}{\displaystyle\sum_{j,k \ne j} A_{jk}}
            \times \left(\frac{\displaystyle\sum_{j,k \ne j} \delta(c_j, c_k)}{n(n - 1)}\right)^{-1}.
\end{align}
As above, $\delta(c_j, c_k) = 1$ if nodes $j$ and $k$ are in the same module and $\delta(c_j, c_k) = 0$ if they are not. The first term in the product, the coverage, measures what fraction of the edges lies within modules, while the second term normalizes by the size of the modules. In other words, $M$ is the ratio of the density of the subgraph formed by the modules over the density of the original graph. Unlike Newman's modularity, maximizing this measure of modularity will not yield meaningful partitions of the network, because it greatly favors small modules. However, if a partition of a network is obtained by other means, such as hierarchical clustering, $M$ quantifies how much the density of the clusters exceeds that of the entire network --- a measure of modular structure.

If additional information is available about a network, the definition of modularity can be adapted to accommodate this information. For example, in the simulation described in section \ref{sec:spontaneousEmergence} a natural partitioning of the network arises from the horizontal gene transfer segments. Since horizontal gene transfer is restricted to predefined blocks, a sensible measure of modularity is given by the number of non-zero entries in the connection matrix within these blocks. For example, if there are 12 blocks of length 10 each, the quantity
\begin{align}
    \sum_{k = 0}^{11} \sum_{j = 1}^{10} \sum_{j' = j + 1}^{10} \Delta_{10k+j,10k+j'},
\end{align}
where $\Delta$ is the connection matrix, quantifies how many of the interactions take place within the predefined blocks. Because the total number of interactions is constant, if this number is large, then more interactions occur within the blocks than between them, indicating a modular structure.

The bandedness in figures \ref{fig:heModRandom} and \ref{fig:heModularity} is a proxy for modularity measuring the locality of interactions in the network. To measure bandedness, the adjacency matrix of the network has to be reordered to concentrate interactions along the diagonal. This can be done using hierarchical clustering \cite{He2009}. The bandedness is then defined as the ratio of the fraction of the interactions within a band along the diagonal over the fraction of the area within the band:
\begin{align}
    \frac{\displaystyle \sum_{0<\vert i-j \vert <W}^D A_{ij}}{\displaystyle \sum_{i \ne j}  A_{ij}}
        \times \left(\frac{\displaystyle \sum_{0<\vert i-j \vert <W} 1}{\displaystyle \sum_{i \ne j} 1}\right)^{-1}
\end{align}
Here, $W$ is the width of the band, and $A_{ij}$ are the elements of the adjacency matrix.

\subsection{Cophenetic Correlation Coefficient}
For a given network one can define a distance between any pair of nodes using, for example, the commute distance described in section \ref{sec:modularityInEcology}. Once distances are defined, one can construct a hierarchical tree, or dendrogram, of the network using hierarchical clustering. A dendrogram can be created for any network regardless of whether it exhibits a hierarchical structure or not. From this dendrogram a new pair-wise distance between nodes can be obtained given by the height at which two nodes are joined in the dendrogram. If the network is hierarchical, the distances obtained from the dendrogram will faithfully represent the original distances, but if the network does not have a hierarchical structure, the pair-wise distances from the dendrogram will differ greatly from the original distances. The condition for a set of pair-wise distances to be tree-like is that for any triple of nodes
\begin{align}
    T_{ij} \le \max\left[T_{ik}, T_{jk}\right],
\end{align}
where $T_{ij}$ is the distance between nodes $i$ and $j$ etc.\ The cophenetic correlation coefficient (CCC) quantifies how well the tree-like representation describes the network from which it was constructed and is thus a quantitative measure of the hierarchy in a network. It is defined as the Pearson correlation coefficient between the node-node distances in the original network and those in the dendrogram:
\begin{align}
   \operatorname{CCC} = \frac{\displaystyle \sum_{i < j} \left(T_{ij} -
T\right)\left(c_{ij} - c\right)}
           {\displaystyle \sqrt{\sum_{i < j} \left(T_{ij} - T\right)^2
\sum_{i < j} \left(c_{ij} - c\right)^2}}.
\end{align}
Here $T$ is the average of the distances in the original network, $T_{ij}$, and $c$ is the average of the dendrogram distances, $c_{ij}$. Unlike measures of modularity which depend on how the network is partitioned, the CCC is a property of the network. It can, however, be affected by the chosen distance measure on the original data and hierarchical clustering algorithm.

  \bibliographystyle{elsarticle-num} 
  \bibliography{coll.bib}





\end{document}